\documentclass{aa}  
 \usepackage{graphicx}
 \usepackage{amssymb}
\usepackage{natbib}
\usepackage{booktabs}
\usepackage{multirow}
\usepackage{pdflscape}

\usepackage{collcell}
\usepackage{xstring}
\usepackage{datatool}
\usepackage{booktabs}
\usepackage{environ}

\usepackage{ulem} 
\usepackage{color}

\offprints{\email{fabrice.mottez@obspm.fr }}

\title{Repeating fast radio bursts caused by small bodies orbiting a pulsar or a magnetar}
\titlerunning{Small bodies, pulsars, and FRB}
\authorrunning{F. Mottez et al.}

\date{\today}
\author{Fabrice Mottez \inst{1}, Philippe Zarka \inst{2}, Guillaume Voisin \inst{3,1} }

   	\institute{LUTH, Observatoire de Paris, PSL Research University, CNRS, Universit\'e de Paris, 
   	5 place Jules Janssen, 92190 Meudon, France 
   	\and 
   	LESIA, Observatoire de Paris, PSL Research University, CNRS, Universit\'e de Paris, Sorbonne Universit\'e,
   	5 place Jules Janssen, 92190 Meudon, France.
    \and
	Jodrell Bank Centre for Astrophysics, Department of Physics and Astronomy, The University of Manchester, Manchester M19 9PL, UK}

\abstract{Asteroids orbiting into the highly magnetized and  highly relativistic wind of a pulsar offer a favourable configuration for repeating fast radio bursts (FRB). The body in direct contact with the wind develops a trail formed of a stationary Alfv\'en wave, called an \textit{Alfv\'en wing}. When an element of wind crosses the Alfv\'en wing, it sees a rotation of the ambient magnetic field that can cause radio-wave instabilities. In 
the observer's
reference frame, the waves are collimated in a very narrow range of directions, and they have an extremely high intensity. A previous work, published in 2014, showed that planets orbiting a pulsar can cause FRB when they pass in our line of sight. We predicted periodic FRB. Since then random FRB repeaters have been discovered.}
	{  We present an upgrade of this theory where repeaters can be explained by the interaction of smaller bodies with a pulsar wind. }
	{ Considering the properties of relativistic Alfv\'en wings attached to a body in the pulsar wind,  and taking thermal consideration into account we conduct a parametric study.} 
	 {{ We find that FRBs, including the Lorimer burst (30 Jy), can be explained by small size pulsar companions (1 to 10 km) between 0.03 and 1 AU from a highly magnetized millisecond pulsar. Some sets of parameters are also compatible with a magnetar. Our model is compatible with the high rotation measure of FRB121102. The bunched timing of the FRBs is the consequence of a moderate wind turbulence. As asteroid belt composed of less than 200 bodies would suffice for the FRB occurrence rate measured with FRB121102.}}
	   { This model, after the present upgrade, is compatible with the properties discovered since its first publication in 2014, when repeating FRB were still unknown. It is based on standard physics, and on common astrophysical objects that can be found in any kind of galaxy. It requires $10^{10}$ times less power than (common) isotropic-emission FRB models. 
   	}

\begin{document}

\maketitle

\section{Introduction}

\citet{Mottez_2014} (hereafter MZ14) proposed a model of FRBs that involves common celestial bodies, neutron stars (NS) and planets orbiting them, well proven laws of physics (electromagnetism),  and a moderate energy demand that allows for a narrowly beamed continuous radio-emission from the source that sporadically illuminates the observer. Putting together these ingredients, the model is compatible with the localization of FRB sources at cosmological distances \citep{Chatterjee_2017}, it can explain the milliseconds burst duration, the flux densities above 1 Jy, and the range of observed frequencies.

{MZ14 is based on the relativistic Alfv\'en Wings theory of \citet{Mottez_2011_AWW}, and concluded that planetary companions of standard or millisecond pulsars could be the source of FRBs.}
In an erratum re-evaluating the magnetic flux and thus the magnetic field in the pulsar wind, \citet{MH20} (hereafter MH20) revised the emitted radio frequency and flux density, and concluded that observed FRB characteristics are rather compatible with companions of millisecond highly magnetized pulsars. These pulsar characteristics correspond to young neutron stars.
The present paper is an upgrade of this revised model, in the light of the discovery of repeating radio bursts made since the date of publication of MZ14  \citep{spitler_repeating_2016,CHIME_2019}. The main purpose of this upgrade is the modeling of the random repeating bursts, and {explanation of} the strong linear polarization possibly associated with huge magnetic rotation measures \citep{Michilli_2018, Gajjar_2018}. 

\begin{figure*}
 	\includegraphics[width=\textwidth]{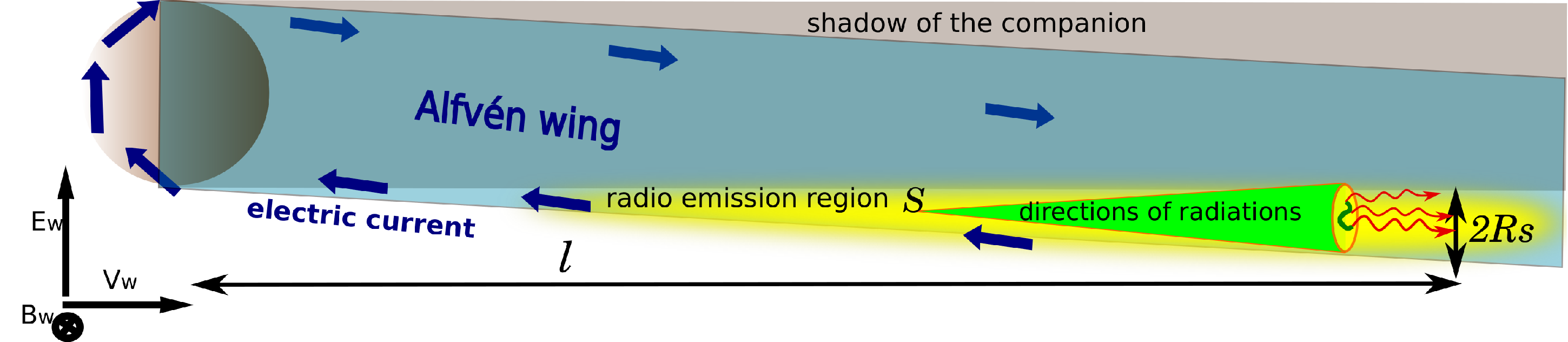}
 	\caption{Schematic view of the 
wake of a pulsar companion. The pulsar is far to the left {out of the figure}. The wind velocity $\vec v_\mathrm{w}$, the wind magnetic field $\vec B_\mathrm{w}$ and the convection electric field $\vec E_\mathrm{w}$ directions are plotted on the left-hand side. The shadow is in gray. The Alfv\'en wing is in blue. The region from where we expect radio emissions (in yellow) is inside the wing and outside the shadow.  {Its largest transverse radius is $R_\mathrm{S}$ and its distance to the companion is $l$. Seen from the NS at a distance $r$, it subtends a solid angle $\Omega_{\sigma}$}. Any point source S in the source region can generate radio waves. Because of the relativistic aberration, their directions are contained in a narrow beam (in green){ of solid angle $\pi \gamma^{-2}$}. If the emission is not isotropic in the source reference frame, it is emitted in a subset of this cone (dark green {tortuous} line) from which emerging photons are marked with red arrows. {Seen from the source point S, this dark green area subtends a solid angle $\Omega_{beam}$.}}  
 	\label{fig_source_size}
 \end{figure*}

The MZ14 model consists of a planet orbiting a pulsar and embedded in its ultra-relativistic and highly magnetized wind. The model requires a direct contact between the wind and the planet. 
Then,  the disturbed plasma flow reacts by creating  a strong potential difference across the companion. This is the source of an electromagnetic wake called Alfv\'en wings (AW) because it is formed of one or two stationary Alfv\'en waves.  AW support an electric current and an associated change of magnetic field direction.
 According to MZ14, when the wind crosses an Alfv\'en wing, it sees a temporary rotation of the magnetic field. This perturbation can be the cause of a plasma instability generating coherent radio waves. Since the pulsar companion and the pulsar wind are permanent structures, this radio-emission process is most probably permanent as well. 
 The source of these radio waves being the pulsar wind when it crosses the Alfv\'en wings, and the wind being highly relativistic with Lorentz factors up to an expected value $\gamma \sim 10^6$, the radio source has a highly relativistic motion relatively to the radio-astronomers who observe it. Because of the relativistic aberration that results, all the energy in the radio waves is concentrated into a narrow beam of aperture angle $\sim 1/\gamma$ rad  (green cone attached to a source S in   Fig. \ref{fig_source_size}). Of course, we observe the waves only when we cross the beam. The motion of the beam (its change of direction) results primarily from the orbital motion of the pulsar companion. The radio-wave energy is evaluated in MZ14 and MH20, as well as its focusing, and it is shown that it is compatible with a brief emission observable at cosmological distances ($\sim$ 1 Gpc) with flux densities larger than 1 Jy. 

In the non-relativistic regime, the Alfv\'en wings of planet/satellite systems and their radio emissions have been well studied, because this is the electromagnetic structure characterizing the interaction of Jupiter and its rotating magnetosphere with its inner satellites Io, Europa and Ganymede \citep{Saur_2004,Hess_2007,Pryor_2011,Louis_2017,Zarka_2018}. It is also observed with the Saturn-Enceladus system \citep{Gurnett_2011}. 

A central question is: ``how can the pulsar wind be in direct contact with the obstacle ?''
{The solution proposed in MZ14 was to assume that the pulsar wind is sub-Alfv\'enic.

In spite of the wind velocity being almost the speed of light $c$, the MZ14 model assumed that 
the wind is slower than Alfv\'en and fast magnetosonic waves. The planet {was then supposed to orbit} inside this sub-Alfvénic. region of the wind.}
{The Alfv\'enic mach number is}
\begin{equation} \label{eq_mach_a}
{M_A}^2 =  \left(\frac{v_r}{c}\right)^2 \left[1 + \frac{\gamma}{\sigma} \left(\frac{v_r}{c}\right)\right],
\end{equation}
{where $v_r$ is the radial wind velocity,  and $\sigma$ is the magnetization parameter, defined as the ratio of magnetic to kinetic energy densities \citep{Mottez_2011_AWW}. Because the Alfv\'en velocity is very close to $c$, and the fast magnetosonic velocity is even closer to $c$ (see \citet{Keppens_2019_disp_relativiste}), the fast magnetosonic Mach number is smaller but close to $M_A$.
However, supposing ${M_F} <1$ or ${M_A} <1$ } implies a low plasma density, difficult to conciliate with current pulsar wind models \citep[e.g.,][]{Timokhin_2015}.

{In this work, however, the necessity of a wind that would be lower than fast magnetosonic waves or than  Alfv\'en  waves  vanishes due to the absence of atmosphere around small bodies.}  A bow shock is created by the interaction of a flow with a gaseous atmosphere, and because asteroids have no atmosphere, there will be no bow shock in front of the asteroid whatever the Mach number of the pulsar wind. Then, the pulsar wind is directly in contact with the surface of the object, which is the only requirement of the Alv\'en wing theory. With super-Alfv\'enic flows, we can allow the wind to have the larger Lorentz factors required to compensate for the smaller size of the companion.
We will {indeed} see in sections \ref{parametrique_pulsars} and \ref{parametrique_magnetars} that  a Lorentz factor $\gamma >5.10^5$ {is} needed for FRBs to be caused by asteroids instead of planets.

 The upgraded model presented here is a generalization of MZ14. In MZ14, it was supposed that the radiation mechanism was the  cyclotron maser instability (CMI), a relativistic wave instability that is efficient in the above mentioned solar-system Alfv\'en wings, with emissions at the local electron gyrofrequency and harmonics \citep{Freund_1983}. Here,{ without entering into the details of the radio emission processes,} we consider that other instabilities as well can generate coherent radio emission, at frequencies much lower than the gyrofrequency,
as is the case in the highly magnetized inner regions of pulsar environments. Removing the constraint that the observed frequencies are above the gyrofrequency allows for FRB radiation sources, i.e. pulsar companions, much closer to the neutron star. 
 Then, it is possible to explore the possibility of sources of FRB such as belts or streams of asteroids at a close distance to the neutron star. Of course, the ability of these small companions to resist evaporation must be questioned. This is why the present paper contains a detailed study of the companions thermal balance.  
 
 
 The model in MZ14 with a single planet leads to the conclusion that FRBs must be periodic, with a period equal to the companion orbital period, provided that propagation effects do not perturb too much the conditions of observation. The possibility of clusters {or belts} of asteroids could explain the existence of non-periodic FRB repeaters, as those already discovered.  
 
 The present paper contains also a more elaborate reflection than in MZ14 concerning the duration of the observed bursts, {and their non periodic repetition rates}. 
 
The MZ14 model proposed a generation mechanism for FRB, leading to predict isolated or periodic FRB. We adapt it here to irregularly repeating FRB.
 Nevertheless, we can already notice that it contains a few elements that have been discussed separately in more recent papers. For instance, the energy involved in this model is smaller by orders of magnitudes than that required by quasi-isotropic emission models, because it involves a very narrow beam of radio emission. This concept alone is discussed in \citet{Katz_wandering_17} where it is concluded, as in MZ14, that a relativistic motion in the source, or of the source, can explain the narrowness of the radio beam. This point is the key to the observability of a neutron star-companion system at cosmological distances.

The MZ14 propose the basic mechanism. We broaden the range of companions.  Using the main results of MZ14 and a thermal analysis summarized in section \ref{section:key_results}, we perform a parametric study and we select relevant solutions (section \ref{sec_parametric}). Compatibility with high rotation measures measured, for instance, with FRB121102, is analyzed.
In section \ref{sec_duration_multiplicity}, we present an analysis of FRB timing.  Before the conclusion, a short discussion is presented, and the number of asteroids required to explain the observed bursts rates is given.

{A table of notations is given in appendix \ref{table_notation}. For energy fluxes, those emitted by the pulsar in all directions are noted with a $L$ (for instance $L_\mathrm{sd}$ is the pulsar spin-down power), and those captured or emitted by the companion are noted with a $\dot E$.}

\section{Pulsar companion in a pulsar wind: key results} \label{section:key_results}

We consider that the radio source is conveyed by the ultra-relativistic pulsar wind. 
The radio emission starts when the pulsar wind crosses the environment of a companion of radius $R_\mathrm{c}$ orbiting the pulsar. 
  
\subsection{Relativistic aberration and solid radiation angle} \label{section_solid_angle}

Let us consider a source of section $\sigma_\mathrm{source}$, at a distance $r$ from the neutron star, with each point of this source emitting radio-waves in a solid angle $\Omega_\mathrm{beam}$ centered on the radial direction connecting it to the neutron star. The solid angle covered by this area as seen from the pulsar is  
$\Omega_\mathrm{\sigma}=\sigma_\mathrm{source}/ r^2$. 
Let us characterize the source by an extent $R_\mathrm{s}$ (fig. \ref{fig_source_size}), then $\sigma_\mathrm{source} = \pi R_\mathrm{s}^2$ and $\Omega_\mathrm{\sigma}=\pi (R_\mathrm{s}/r)^2$.  Following MZ14, the source is attached to the pulsar wind during the crossing of the companion's environment. In the reference frame of the source, we suppose that the radio emission is contained in a solid angle $\Omega_\mathrm{A}$
that can be $\ll 4\pi$.
Because of the relativistic aberration, the radiation is emitted in a solid angle 
\begin{equation}
\label{eq:omegabeam}
\Omega_\mathrm{beam}= \pi \gamma^{-2} (\Omega_\mathrm{A}/4 \pi).
\end{equation}
The total solid angle of the radio emissions in the observer's frame is $\Omega_\mathrm{T}>\Omega_\mathrm{\sigma}+\Omega_\mathrm{beam}= \pi (s/r)^2 +  \pi \gamma^{-2} (\Omega_\mathrm{A}/4 \pi)$. The parametric study conducted in section \ref{parametrique_pulsars} shows that $\gamma \sim 10^6$ and $s<500$ m and $r>0.01$ AU. Then, practically,  $\Omega_\mathrm{T} \sim\Omega_\mathrm{beam}= \gamma^{-2} (\Omega_\mathrm{A}/4)$. This is the value used in the rest of the paper.
The strong relativistic aberration is a factor favoring the possibility of observing a moderately intense phenomenon over cosmological distances, when the very narrow radiation beam is crossed. 

\subsection{Alfv\'en wings} \label{section_AW_general}

We consider a pulsar companion in the ultra-relativistic wind of the pulsar. The wind velocity modulus is $V_\mathrm{W} \sim c$.
The magnetic field in the inner magnetosphere (distance to the neutron star $r < r_\mathrm{LC}$ where $r_\mathrm{LC}$ is the light cylinder radius) is approximately a dipole field. In the wind, its amplitude dominated by the toroidal component decreases as $r^{-1}$, and the transition near $r_\mathrm{LC}$ is continuous,
\begin{eqnarray} \nonumber
B&=& B_* \left(\frac{R_*}{r}\right)^3  \mbox{ for } r < r_\mathrm{LC},\\ \label{eq_B_dipole}
B&=& B_* \left(\frac{R_*^3}{r_\mathrm{LC}^2 r}\right) \mbox{ for } r \ge r_\mathrm{LC}.
\end{eqnarray}

The orbiting companion is in direct contact with the pulsar wind. 
The wind velocity relatively to the orbiting companion $\vec{V}_\mathrm{w}$ crossed with the ambient magnetic field $\vec{B}_\mathrm{w}$ is the cause of an induced electric structure associated with an average field $\vec{E_0}=\vec{V}_\mathrm{w} \times \vec{B}_\mathrm{w}$, called a unipolar inductor \citep{Goldreich_Lynden_Bell_1969}. 

The interaction of the unipolar inductor with the conducting plasma generates a stationary Alfv\'en wave attached to the body, called Alfv\'en wing \citep{Neubauer_1980}. The theory of Alfv\'en wings has been revisited in the context of special relativity by \cite{Mottez_2011_AWW}. 
In a relativistic plasma flow, the current system carried by each AW (blue arrows in fig. \ref{fig_source_size}), ``closed at infinity'', has an intensity $I_\mathrm{A}$ related to the
AW power $\dot E_\mathrm{A}$ by
\begin{equation} \label{courant_puissance_AW}
\dot E_\mathrm{A} =I_\mathrm{A}^2 \mu_0 c,
\end{equation}
where $\mu_0 c =1/377$ mho is the vacuum conductivity. 
From  \citet{Mottez_2014} and MH20, the computation of the current $I_\mathrm{A}$  in the relativistic AW combined with the characteristics of the pulsar wind provides an estimate of the AW power,
\begin{equation} \label{puissance_AW}
\dot E_\mathrm{A} = \frac{\pi}{\mu_0 c^3} {R_c}^2{r}^{-2} R_*^6 B_*^2 \Omega_*^4.
\end{equation}
(See Table \ref{table_notation} for notations.)

\subsection{Alfv\'en wing radio emission} \label{section_AW}

By extrapolation of known astrophysical systems, it was shown in MZ14 that the Alfv\'en wing is a source of radio-emissions of power 
\begin{equation} \label{puissance_radio}
\dot E_\mathrm{R} =\epsilon \dot E_\mathrm{A}
\end{equation}
where $2.\,10^{-3} \le \epsilon \le 10^{-2}$ \citep{Zarka_2001,Zarka_2007}. The resulting flux density at distance $D$ from the source is 
\begin{eqnarray} \label{S_observe} \nonumber
\left(\frac{S}{\mbox{Jy}}\right) = 10^{-27} A_{\mathrm{cone}}&&\left(\frac{\epsilon}{10^{-3}}\right) \left( \frac{\gamma}{10^5}\right)^2  \left( \frac{\dot E_\mathrm{A}}{W} \right)  \times \\
&&\left( \frac{\mbox{Gpc}}{D} \right)^2 \left(\frac{\mbox{1 GHz}}{\Delta f}\right).
\end{eqnarray}
where $\Delta f$ is the spectral bandwidth of the emission, $\gamma$ is the pulsar wind Lorentz factor, and $\gamma^2$ in this expression is a consequence of the relativistic beaming: the radio emissions are focused into a cone of characteristic angle $\sim \gamma^{-1}$ when $\gamma \gg 1$.

The coefficient $A_{\mathrm{cone}}$ is an anisotropy factor. Let $\Omega_\mathrm{A}$ be the solid angle in which the radio-waves are emitted in the source frame. Then, $A_{\mathrm{cone}}=4 \pi/\Omega_\mathrm{A}$. If the radiation is isotropic in the source frame, $A_{\mathrm{cone}}=1$, otherwise, $A_{\mathrm{cone}}>1$. For instance, with the CMI, $A_{\mathrm{cone}}$ up to 100
\citep{Louis_2019}.

{Since the wing is powered by the pulsar wind, the observed flux density of equation (\ref{S_observe})
	can be related to the properties of the pulsar by expressing $\dot E_\mathrm{A}$ as a function of $P_*$ and $B_*$, respectively the spin period and surface magnetic field of the neutron star, and the size of the object $R_\mathrm{c}$.}
Thus, equation (\ref{S_observe}) becomes,
\begin{eqnarray} \nonumber
\left(\frac{S}{\mbox{Jy}}\right) &=& 2.7 \times 10^{-9} A_{\mathrm{cone}} \left( \frac{\gamma}{10^5}\right)^2 \left(\frac{\epsilon}{10^{-3}}\right) \left( \frac{R_\mathrm{c}}{10^7 \mbox{m}} \right)^2 \times \\ \nonumber 
& &\left( \frac{1 \mbox{AU}}{r} \right)^{2} \left( \frac{R_*}{10^4 \mbox{m}} \right)^6 \times 
\\ \label{eq_flux_density_reduced}
& & \left( \frac{B_*}{10^5 \mbox{T}} \right)^2 \left(\frac{10 \mbox{ms}}{P_{*}}\right)^4 \left( \frac{\mbox{Gpc}}{D} \right)^2 \left(\frac{\mbox{1 GHz}}{\Delta f}\right).
\end{eqnarray}

{Another way of presenting this result is in terms of equivalent isotropic luminosity $\dot E_\mathrm{iso,S}=4 \pi S D^2 \Delta f$, which in terms of reduced units reads}
\begin{equation}\label{EisoS}
\left( \frac{\dot E_\mathrm{iso,S}}{\mbox{W}}\right)= 1.28 \times 10^{35} \left( \frac{S}{\mbox{Jy}}\right)\left( \frac{D}{\mbox{Gpc}}\right)^2 \left( \frac{\Delta f}{\mbox{GHz}}\right).
\end{equation}
Observations of FRB of known distances (up to $\sim$ 1 Gpc) reveal values of $\dot E_\mathrm{iso,S}$ in the range $10^{33}$ to $10^{37}$ W \citep{Luo_2020}. This is compatible with  fiduciary values appearing in  Eq. (\ref{EisoS}). 

{Another way of computing the radio flux is presented in Appendix \ref{section_AW_bis}. It is found that}
\begin{eqnarray} \nonumber
\left( \frac{\dot E'_\mathrm{iso,S}}{\mbox{W}}\right)= 3.2 \times 10^{29} \left( \frac{R_*}{10\mbox{ km}}\right)^6  \left( \frac{B_*}{10^5 \mbox{T}}\right)^2  \times \\
\left( \frac{10\mbox{ ms}}{P_*}\right)^4 \left( \frac{R_c}{10^4\mbox{km}} \right)^2 \left( \frac{AU}{\mbox{r}}\right)^2 \left( \frac{\gamma}{10^5}\right)^2. \label{E_iso_prime}
\end{eqnarray}
where the prime denotes the alternate way of computation. 
In spite of being introduced with different concepts, without any explicit mention of Alfv\'en wings, Eq. (\ref{E_iso_prime}) and Eq. (\ref{EisoS}) should be compatible. In the parametric study, we will check their expected similarity
\begin{equation}\label{check_isotropes}
\dot E_\mathrm{iso,S}' \sim \dot E_\mathrm{iso,S}.
\end{equation}

\subsection{Frequencies} \label{sec_freq}

It was proposed in MZ14 that the instability triggering the radio emissions is the cyclotron maser instability. Its low-frequency cutoff is the electron gyrofrequency. 
We will see that for FRB121102 and other repeaters, {the CMI cannot always explain the observed frequencies in the context of our model}. 

In any case, we still use the electron gyrofrequency $f_\mathrm{ce,o}$ in the observer's frame as a useful scale.
{From MZ14, and MH20} 
\begin{eqnarray} \nonumber
\left(\frac{f_\mathrm{ce,o}}{\mbox{Hz}}\right)&=& 5.2 \times 10^4 \left(\frac{\gamma}{10^5}\right)
\left(\frac{B_{*}}{10^5 \mbox{T}} \right)
\times 
\\ \nonumber
& &\left(\frac{1 \mbox{AU}}{r}\right)^{2} \left( \frac{R_*}{10^4 \mbox{m}} \right)^{3}  \left(\frac{10 \mbox{ms}}{P_{*}}\right)\times
\\
& &\left\{1+ \left[\frac{\pi \, 10^5}{\gamma} \left(\frac{10 \mbox{ms}}{P_{*}}\right)\left(\frac{r}{1 \mbox{AU}}\right) \right]^2\right\}^{1/2}.
\label{eq_omega_observer}
\end{eqnarray}

\subsection{Pulsar spin-down age}
The chances of observing a FRB triggered by asteroids orbiting a pulsar will be proportional to the pulsar spin-down age
$\tau_\mathrm{sd}= P_*/2 \dot P_*$. This definition combined with $L_\mathrm{sd} = - I \Omega_* \dot \Omega_*$ gives an expression depending on the parameters of our parametric study, $\tau_\mathrm{sd}={2 \pi^2 I}{P_*^{-2} L_\mathrm{sd}^{-1}} $. 
In reduced units, 
\begin{equation}\label{Tdynamique}
\left(\frac{\tau_\mathrm{sd}}{\mbox{yr}}\right) =  6.6 \, 10^{10} \left(\frac{I}{10^{38}\mbox{ kg.m}^{2}}\right) \left(\frac{\mbox{10 ms}}{P_*}\right)^2 \left(\frac{10^{25}\mbox{ W}}{L_\mathrm{sd}}\right)
\end{equation}
We can check, for consistency with observational facts, that a Crab-like pulsar ($P_*=33$ ms and $L_\mathrm{sd}=10^{31}$ W) has a spin-down age $\tau_\mathrm{sd} \sim 6600$ yr, that is compatible with its age since the supernova explosion in 1054.

\subsection{Minimal asteroid size required against evaporation} \label{section_thermique}

The thermal equilibrium temperature $T_\mathrm{c}$ of the pulsar companion is 
\begin{equation} \label{puissance_temperature}
\dot E_\mathrm{T}+ \dot E_\mathrm{P}+ \dot E_\mathrm{NT}+ \dot E_\mathrm{W}+ \dot E_\mathrm{J} = 4 \pi \sigma_\mathrm{S} R_\mathrm{c}^2 T_\mathrm{c}^4,
\end{equation}
{where $\sigma_\mathrm{S}$ is the Stefan-Boltzmann constant, and $R_\mathrm{c}$ and $T_\mathrm{c}$ are radius and temperature of the companion  respectively.}
The source $\dot E_\mathrm{T}$ is the black-body radiation of the neutron star, $\dot E_\mathrm{P}$ is associated with inductive absorption of the Poynting flux, $\dot E_\mathrm{NT}$ is caused by the pulsar non-thermal photons, $\dot E_\mathrm{W}$ is associated with the impact of wind particles, and $\dot E_\mathrm{J}$ is caused by the circulation of the AW current $I_\mathrm{A}$ into the companion.  

{These sources of heat are investigated in appendix \ref{section_heat_sources}.} 
In the appendix, we introduce two factors $f$ and $g$ used to abridge the sum $\dot E_\mathrm{NT}+\dot E_\mathrm{W}=(1-f) g L_\mathrm{sd}$, into a single number which is kept in the parametric study.

The pulsar companion can survive only if it does not evaporate. As we will see in the parametric studies, the present model works better with metallic companions. Therefore, we anticipate this result and we consider that its temperature must not exceed the iron fusion temperature $T_{\mathrm{max}} \sim 1400$ K. 
Inserting $T_{\mathrm{max}}$ in equation (\ref{puissance_temperature})
provides an upper value of $\dot E_\mathrm{A}$, 
whereas Eq. (\ref{S_observe}) with a minimum value $S = 1$ Jy provides a lower limit $\dot E_\mathrm{A min}$ of $\dot E_\mathrm{A}$. 
The combination of these relations sets a constrain on the companion radius
\begin{eqnarray}  \label{contrainte_R_c_0}
R_\mathrm{c}^3 \mu_0 c \sigma_\mathrm{c} \left[ 4 \pi \sigma_\mathrm{S} T_{\mathrm{max}}^4 
-\frac{X}{4 r^2}  \right] >  \dot E_\mathrm{A min}
\end{eqnarray}
where $E_\mathrm{A min}$ is given by Eq. (\ref{S_observe}) with $S =1$ Jy, 
and 
\begin{equation}
X=  4 \pi R_*^2 \sigma_\mathrm{S} T_*^4+ ((1-f)  g+ \frac{f}{f_\mathrm{p}} Q_\mathrm{{abs}}) L_\mathrm{sd}  
\end{equation}
 Because the coefficient of inductive energy absorption $Q_\mathrm{{abs}} < Q_\mathrm{{max}}= 10^{-6}$, and $f_\mathrm{p}$ is a larger fraction of unity than $f$
 we can neglect the contribution of the inductive heating $\dot E_\mathrm{P}$. 
Combining with Eq. (\ref{S_observe}), with $T_{\mathrm{max}}=1400$ K for iron fusion temperature, and using normalized figures,
we finally get the condition for the companion to remain solid:
\begin{eqnarray} \label{eq_XY}
R_\mathrm{c}^3 \left(\frac{\sigma_\mathrm{c}}{10^7 \mathrm{mho}}\right)
\left[2.7 \, 10^4 \left(\frac{T_\mathrm{c}}{1400 \mathrm{K}}\right)^4 -\left(\frac{\mathrm{AU}}{r}\right)^2 X \right]
\\ \nonumber 
>2.7 \, { 10^{15} } \frac{Y}{A_{\mathrm{cone}}}
\end{eqnarray}
where 
\begin{equation} \label{eq_X}
X=7 \left(\frac{R_*}{10^4 \mathrm{m}}\right)^2
\left(\frac{T_*}{10^6 \mathrm{K}}\right)^4
+ (1-f) g
\left(\frac{L_\mathrm{sd}}{10^{25} \mathrm{W}}\right)
\end{equation}
and 
\begin{equation} \label{eq_Y}
Y=\left(\frac{S_\mathrm{min}}{\mathrm{Jy}}\right)
{ \left(\frac{10^5}{\gamma}\right)^2}
\left(\frac{10^{-3}}{\epsilon}\right)
\left(\frac{D}{\mathrm{Gpc}}\right)^2
\left(\frac{\Delta f}{\mathrm{GHz}}\right).
\end{equation}
We note that when, for a given distance $r$, the factor of $R_\mathrm{c}^3$ in Eq. (\ref{contrainte_R_c_0}) is negative,
then no FRB emitting object orbiting the pulsar can remain in solid state, whatever its radius. 

\subsection{Clusters and belts of small bodies orbiting a pulsar} \label{sec_clusters_belts}

With FRB121102, no periodicity was found in the time distribution of bursts arrival. Therefore, we cannot consider that they are caused by a single body orbiting a pulsar. 
\citet{Scholz_2016} have shown that the time distribution of bursts of FRB121102 is clustered. Many radio-surveys lasting more than 1000 s each and totaling 70 hours showed no pulse occurrence, while 6 bursts were found within a 10 min period. The arrival time distribution is clearly highly non-Poissonian. {This is somewhat at odds with FRB models based on pulsar giant pulses which tend to be Poissonian \citep{karuppusamy_giant_2010}.}

We suggest that the clustered distribution of repeating FRBs could result from 
asteroid swarms.
{The hypothesis of clustered asteroids could be confirmed by the recent detection of
a 16.35 $\pm$ 0.18 day periodicity of the pulses rate of the repeating FRB 180916.J0158+65 detected by
the Canadian Hydrogen Intensity Mapping Experiment Fast Radio Burst Project
(CHIME/FRB). 
Some cycles show no bursts, and some show multiple bursts \citep{chime_periodic_2020}.}

{Other irregular sources could be considered.}
We know that planetary systems exist around pulsar, if uncommon. One of them has four planets. It could as well be a system formed of a massive planet and Trojan asteroids. In the solar system, Jupiter is accompanied by
about $1.6 \times 10^5$  Trojan asteroids bigger than 1 km radius near the L4 point of the Sun-Jupiter system
\citep{Jewitt2000}. Some of them reach larger scales: 588 Achilles measures about 135 kilometers in diameter, and 617 Patroclus, 140 km in diameter, is double.  The Trojan satellites have a large distribution of eccentric orbits around the Lagrange points  \citep{Jewitt_2004}. If they were observed only when they are rigorously aligned with, say, the Sun and Earth, the times of these passages would look quite non-periodic, non-Poissonian, and most probably clustered.

\citet{Gillon_2017} have shown that rich planetary systems with short period planets  are possible around main sequence stars. There are seven Earth-sized planets in the case of Trappist-1, the farther one having an orbital period of only 20 days. Let us imagine a similar system with planets surrounded by satellites. The times of alignment of satellites along a given direction would also seem quite irregular, non-Poissonian and clustered.

{Also very interesting is the plausible detection of an asteroid belt around PSR B1937+21 \citep{shannon_asteroid_2013} which effect is detected as timing red noise.
It is reasonable to believe that there exists other pulsars where such belts might be presently  not distinguishable from other sources of red noise (see e.g. \citet{shannon_limitations_2014} and references therein).} So, we think that is it reasonable to consider swarms of small bodies orbiting a pulsar, even if they are not yet considered as common according to  current representations.

{In conclusion of this section, the random character of the FRB timing could be due to a large number of asteroids or planets and satellites. These bodies could be clustered along their orbits, causing non-Poissonian FRB clustered FRB timing.  In section \ref{sec_discussion_multiplicity}, a different cause of clustered FRB occurrence is discussed : every single asteroid could be seen several times over a time interval ranging over tens of minutes.}

\section{Parametric study} \label{sec_parametric}
\subsection{Minimal requirements} \label{sec_minimal_requirements}
We wanted to test if medium and small solid bodies such as asteroids can produce FRBs. 
Thus we conducted a few parametric studies based on sets of parameters compatible with neutron stars and their environment. They are based on the equations (\ref{eq_flux_density_reduced},\ref{eq_omega_observer},\ref{eq_XY},\ref{eq_X},\ref{eq_Y})   presented above. 
{We then selected the cases that meet the following conditions: }
(1) the observed signal amplitude on Earth must exceed a minimum value $S_\mathrm{min}=0.3$ Jy, (2) the companion must be in solid state with no melting/evaporation happening, and (3) the radius of the source must exceed the maximum local Larmor radius. This last condition is a condition of validity of the MHD equations that support the theory of Alfv\'en wings \citep{Mottez_2011_AWW}.{ Practically, the larger Larmor radius might be associated with electron and positrons. In our analysis, this radius} is compiled for hydrogen ions at the speed of light, so condition (3) is checked conservatively. 

{Because the Roche limit for a companion density $\rho_\mathrm{c} \sim 3000$ kg.m$^{-3}$ is about 0.01 AU, we do not search for solutions for lower neutron star-asteroid separations $r$.}
{This corresponds to orbital periods longer than 0.3 day for a neutron star mass of 1.4 solar mass.}

\subsection{Pulsars and small-size companions} \label{parametrique_pulsars}

We first chose magnetic fields and rotation periods relevant to standard and recycled pulsars. But it appeared that standard and recycled pulsars do not produce FRB meeting the minimal requirements defined in section \ref{sec_parametric}. Therefore, we extended our exploration to the combinations of the parameters listed in Table \ref{table_parametrique_2}. 
With $\epsilon=2. \, 10^{-3}$, within the 5,346,000 tested sets of parameters, 400,031 (7.5 \%) fulfill the minimum requirements. With the larger radio yield $\epsilon=10^{-2}$, a larger number of 521,797 sets (9.8 \%) fulfill these conditions. According to the present model, they are appropriate for FRB production. 

{Since some of these solutions are more physically meaningful than others, we discuss a selection of realistic solutions. }

{A selection of representative examples of parameter sets meeting the above 3 requirements are displayed in table \ref{table_jeux_de_parametres_Rc_petit}.}

\begin{table*}
	\caption{Parameter set of the first parametric study of FRBs produced by pulsar companions of medium and small size. The last column is the number of values tested for each parameter. The total number of cases tested is the product of all values in the last column, i.e. {$10 692 000$}.} 
	\label{table_parametrique_2} 
	\centering 
	\begin{tabular}{|p{4cm}|p{1.7cm}|l|p{1.2cm}|p{1.5cm}|} 
		\hline 
		Input parameters		& Notation 	& Values 									&	Unit		& Number of values	\\
		\hline
		NS magnetic field 		& $B_*$		& $10^{7+n/2}, n \in\{0,4\}$	 				&	T	 	&	5	\\		
		NS radius				& $R_*$		&  $10,  11, 12, 13$ 							&	km	 	&	4	\\
		NS rotation period		& $P_*$		& $0.003, 0.01, 0.032, 0.1 , 0.32$ 				&	s	 	&	5	\\
		NS temperature		& $T_*$		& $ 2.5 \, 10^5, 5. \, 10^5, 10^6$				&	K		&	3	\\
		Wind Lorentz factor		& $\gamma$	&$3. \, 10^5, 10^6, 3. \, 10^6$					&		 	&	3	\\
		Companion radius		& $R_\mathrm{c}$		& $1, 2.2, 4.6, 10, 22, 46, 100,  316, 1000, 3162, 10000$ &	km	&	11	\\
		Orbital period			& $T_\mathrm{orb}$	& $0.3 \times 2^{n} \;  n \in \{0,11\}$	(from 0.3 to 614) &	day		&	12	\\
		Companion conductivity	& $\sigma_\mathrm{c}$	& $10^{-3}, 10^{2}, 10^7 $ 					&	mho	 	&	3	\\
		Power input			& $(1-f) gL_\mathrm{sd}$& $10^{25}, 10^{26}, 10^{27}, 10^{28}, 10^{29}$ 	& W	&	5	\\
		Radio efficiency		& $\epsilon$	& $2. \, 10^{-3}, 10^{-2}$						& 	 	  	&	2	\\
		Emission solid angle 	& $\Omega_\mathrm{A}$	& $0.1 , 1, 10$     							&	sr	 	&	3	\\
		Distance to observer		& $D$		&$1$ 									&	Gpc	 	&	1	\\
		FRB Bandwidth			& $\Delta f$	& max($1, f_\mathrm{ce}/10$) 						&	GHz	 	&	1	\\
		FRB duration			& $\tau$		& $5. \, 10^{-3}$ 							&	s	 	&	1	\\
		\hline
	\end{tabular}
\end{table*}

We considered a companion radius between 1 and 10,000 km. However, as the known repeating FRBs have high and irregular repetition rates, they should more probably be associated with small and  numerous asteroids. 
For instance, the volume of a single 100 km body is equivalent to those of $10^3$ asteroids of size $R_\mathrm{c}=10$ km. So we restricted the above parametric study to asteroids of size $R_\mathrm{c} \le 10$ km. We also considered energy inputs $(1-f) g L_\mathrm{sd} > 10^{27}$ W, because smaller values would not be realistic for short period and high magnetic field pulsars. (See sections \ref{section_thermique} and \ref{section_wind_and_non_thermal} for the definition of $(1-f) g L_\mathrm{sd}$.)
We also required a pulsar spin-down age $\tau_\mathrm{sd} \ge 10$ yr, and a neutron star radius $R_* \le 12$ km. We found 115 sets of parameters that would allow for observable FRB at 1 Gpc. The cases 3-13 in Table \ref{table_jeux_de_parametres_Rc_petit} are taken from this subset of solutions.

Here are a few comments {on Table \ref{table_jeux_de_parametres_Rc_petit}}.  
The examination of the spin-down age $\tau_\mathrm{sd}$ shows that the pulsars allowing FRB with small companions (cases 1-
{13, companion radius $\le$10 km except case 2 {with} 46 km)}
last generally less than a century (with a few exceptions, such as case 3). 
Most of the parameter sets involve a millisecond pulsar ($P_*$ = 3 or 10 ms, up to 32 msec for cases 1-2), with a Crab-like or larger magnetic field ($B_* \ge 3.\, 10^7$ T). Therefore, only very young pulsars could allow for FRB with asteroids (still with the exception displayed on case 3).
{The required Lorentz factor is $\gamma \ge 10^6$.}
{Constraints on the pulsar radius ($R_* $ down to 10 km) and} temperature (tested up to $10^6$ K) are not strong.
Orbital distances in the range $0.01-0.63$ AU are found.
A power input  $(1-f) gL_\mathrm{sd}$ up to the maximum tested value of $10^{29}$ W
{is found, and with $T_*$ up to $10^6$ K, this does not raise any significant problem for asteroid survival down to distances of 0.1 AU.}
The companion conductivity is 
{found to lie between $10^2$ and $10^7$ mho, implying}
that the companion must be metal rich, or that it must contain some metal. A metal-free silicate or carbon  body could not explain the observed FRB associated with small companion sizes.  
If the duration of the bursts was caused only by the source size, this size would be comprised between 0.21 and 0.53 km. This is compatible with 
{even the smallest companion sizes (1-10 km).
The electron cyclotron frequency varies in the range 40 GHz $\le f_{ce} \le$ 28 THz, always above observed FRB frequencies}.
Therefore, the radio-emission according to the present model cannot be the consequence of the CMI. 
The maximal ion Larmor radius, always less than a few meters, is much smaller than the source size, confirming that there is no problem with MHD from which the Alfv\'en wing theory derives. 

We {find} that even a 2 km asteroid can cause FRBs (case 4). 
Most favorable cases for FRBs with small asteroids ($R_\mathrm{c}=10$ km or less) are unsurprisingly associated with a radio-emission efficiency $\epsilon=10^{-2}$. Nevertheless, the lower value $\epsilon=2. \, 10^{-3}$ still allow for solutions for $r=0.1$ AU (cases 12-13).  
Two sets of parameters {for small asteroids provide a flux density} $S > 20$ Jy (such as case 8). 
Larger asteroids ($R_\mathrm{c}=100$ km) closer to the neutron star ($r=0.025$ AU) could provide a $S=71$ Jy burst, larger than the Lorimer burst (case 14).
We also consider the case of a 316 km asteroid, at the same distance $r=0.025 AU$, which could provide a 7 kJy burst (15*). Nevertheless, we have put an asterisk after this case number, because problematic effects enter into action : (1) the size of the asteroid is equal to the pulsar wavelength, so it must be heated by the pulsar wave Poynting flux, and it is probably undergoing evaporation ; (2) it can also tidally disrupted since it is not far from the Roche limit, and the self-gravitational forces for an {asteroid} of this size are important for its cohesion. 

{The 16.35 day periodicity of the repeating FRB 180916.J0158+65 \citep{chime_periodic_2020}
would correspond, for a neutron star mass of 1.4 solar mass, to a swarm of asteroids at a distance $r=0.14$ AU. This fits the range of values from our parametric study. 
The case of FRB 180916.J0158+65 will be studied in more details in a forthcoming study where emphasis will be put on dynamical aspects of clustered asteroids.} 

Therefore, we can conclude that FRB can be triggered by 1-10 km sized asteroids orbiting a young millisecond pulsar, with magnetic field $B_*$ comparable to that of the Crab pulsar or less, but with shorter period. These pulsars could keep the ability to trigger FRB visible at 1 Gpc during times $\tau_\mathrm{sd}$ from less than a year up to a few centuries. 
{Our results also provide}
evidence of validity of the pulsar/companion model as an explanation for repeating FRBs.

\begin{table*}
	\caption{Examples illustrating the results of the parametric studies in Table \ref{table_parametrique_2} for pulsars with small companions. Input parameters in upper lines (above double line). We abbreviate the maximal admissible value of $(1-f) g L_\mathrm{sd}$ as $L_\mathrm{nt}$. } 
	\label{table_jeux_de_parametres_Rc_petit} 
	\centering 
\begin{tabular}{|c|c|rrrlrccrrrrrr|} 
\hline
\textbf{Parameter}	& case	& $B_*$	& $R_*$	& $P_*$	& $r$	& $R_\mathrm{c}$	& $L_\mathrm{nt}$	& $\sigma_\mathrm{c}$	& $f_\mathrm{ceo}$	& $S$	& $L_\mathrm{sd}$	& $\dot E_\mathrm{iso, A}$	& $\dot E_\mathrm{iso, S}$	& $\tau_\mathrm{sd}$	\\
\hline
\textbf{Unit}		& \#		& T		& km		& ms		& AU		& km		& W		& mho		& GHz		& Jy		& W			& W				& W				& yr			\\	 \hline
\hline \multicolumn{15}{|l|}{\textbf{Pulsar and asteroids}}\\
\hline \multicolumn{15}{|l|}{$\epsilon=10^{-2}$, $\gamma=3.\,10^6$, $\Omega_\mathrm{A}= 0.1$ sr, $T_*=5.\,10^5$ K}\\ \hline 
long $P_*$		& 1		& $3.2\,10^8$ & $12$ & $32$	& $0.01$	& $10$	& $10^{25}$	& $10^2$	& $28003$	& $0.7$	& $1.7\,10^{32}$ & $9.0\,10^{34}$	& $8.6\,10^{34}$	& $381$	\\ 
long $P_*$		& 2		& $10^9$	& $10$	& $32$	& $0.06$	& $46$	& $10^{27}$	& $10^7$	& $1271$		& $1.2$	& $5.8\,10^{32}$ & $1.6\,10^{35}$	& $1.5\,10^{35}$	& $114$	\\ 
\hline \multicolumn{15}{|l|}{$\epsilon=10^{-2}$, $\gamma=3.\,10^6$, $\Omega_\mathrm{A}= 0.1$ sr, $T_*=10^6$ K}\\ \hline 
low $B_*,\sigma_\mathrm{c}$	& 3		& $3.2\,10^7$ & $11$ & $3$	& $0.1$	& $10$	& $10^{27}$	& $10^2$	& $212$		& $0.4$	& $1.0\,10^{34}$ & $5.2\,10^{34}$	& $5.0\,10^{34}$	& $642$	\\ 
small $R_\mathrm{c}$		& 4		& $3.2\,10^8$ & $10$ & $3$	& $0.1$	& $2$	& $10^{27}$	& $10^7$	& $1595$		& $1.0$	& $5.8\,10^{35}$ & $1.4\,10^{35}$	& $1.3\,10^{35}$	& $11$	\\ 
mild    		& 5		& $3.2\,10^8$ & $12$ & $10$	& $0.1$	& $10$	& $10^{27}$	& $10^2$	& $872$		& $0.7$	& $1.7\,10^{34}$ & $8.8\,10^{34}$	& $8.8\,10^{34}$	& $38$	\\ 
large $L_\mathrm{nt}$		& 6		& $10^8$        & $10$ & $3$	& $0.25$	& $ 10$	& $10^{28}$	& $10^7$	& $80$		& $0.3$	& $5.8\,10^{34}$ & $4.6\,10^{34}$	& $4.5\,10^{34}$	& $114$	\\ 
large $L_\mathrm{nt}$		& 7		& $3.2\,10^8$ & $10$ & $3$	& $0.63$	& $ 10$	& $10^{29}$	& $10^7$	& $40$		& $0.5$	& $5.8\,10^{35}$ & $7.3\,10^{34}$	& $7.0\,10^{34}$	& $11$	\\ 
large $S$			& 8		& $3.2\,10^8$ & $10$ & $3$	& $0.1$	& $ 10$	& $10^{27}$	& $10^2$	& $1595$		& $22.$	& $5.8\,10^{35}$ & $3.0\,10^{36}$	& $2.8\,10^{36}$	& $11$	\\ 
large $r$			& 9		& $10^8$        & $12$ & $3$	& $0.4$	& $ 10$	& $10^{27}$	& $10^7$	& $54$		& $0.4$	& $1.7\,10^{35}$ & $5.5\,10^{34}$	& $5.3\,10^{34}$	& $38$	\\ 
\hline \multicolumn{15}{|l|}{$\epsilon=10^{-2}$, $\gamma=3.\,10^6$, $\Omega_\mathrm{A}= 1$ sr, $T_*=10^6$ K}\\ \hline
large $r$			& 10		& $3.2\,10^8$ & $10$ & $3$	& $0.16$	& $ 10$	& $10^{27}$	& $10^2$	& $633$		& $0.9$	& $5.8\,10^{35}$ & $1.2\,10^{35}$	& $1.1\,10^{35}$	& $11$	\\ 
low $B_*$			& 11		& $10^8$        & $11$ & $3$	& $0.1$	& $ 10$	& $10^{27}$	& $10^7$	& $671$		& $0.4$	& $1.0\,10^{35}$ & $5.2\,10^{34}$	& $5.0\,10^{34}$	& $64$	\\ 
\hline \multicolumn{15}{|l|}{$\epsilon=2. \,10^{-3}$,  $\gamma=10^6$, $\Omega_\mathrm{A}= 0.1$ sr, $T_*=10^6$ K}\\ \hline
low $\gamma$		& 12		& $3.2\,10^8$ & $10$ & $3$	& $0.1$	& $ 10$	& $10^{27}$	& $10^7$	& $532$		& $0.5$	& $5.8\,10^{35}$ & $6.6\,10^{34}$	& $6.3\,10^{34}$	& $11$	\\ 
\hline \multicolumn{15}{|l|}{$\epsilon=2. \,10^{-3}$, $\gamma=3.\,10^6$, $\Omega_\mathrm{A}= 0.1$ sr, $T_*=10^6$ K}\\ \hline 
long $\tau_\mathrm{sd}$		& 13		& $10^8 $       & $10$ & $3$	& $0.1$	& $ 10$	& $10^{27}$	& $10^7$	& $504$		& $0.4$	& $5.8\,10^{34}$ & $5.9\,10^{34}$	& $5.7\,10^{34}$	& $114$	\\ 
large $S, \tau_\mathrm{sd}$	& 14		& $3.2\,10^7$ & $10$ & $3$	& $0.025$	& $ 100$	& $10^{25}$	& $10^7$	& $2552$		& $71.$	& $5.8\,10^{33}$ & $9.4\,10^{36}$	& $9.0\,10^{36}$	& $1138$	\\ 
huge $S$			& 15*	& $10^8$        & $10$ & $3$	& $0.025$	& $ 316$	& $10^{25}$	& $10^2$	& $8071$		& $7.\,10^3$ & $5.8\,10^{34}$ &$9.4\,10^{38}$	& $9.0\,10^{38}$	& $114$	\\ \hline  
\hline \multicolumn{15}{|l|}{\textbf{Magnetar and asteroids}}\\ 
\hline \multicolumn{15}{|l|}{$\epsilon=10^{-2}$, $\gamma=3.\,10^6$, $\Omega_\mathrm{A}= 0.1$ sr, $T_*=2. \,10^6$ K}\\ \hline 
small $R_\mathrm{c}$		&16		& $10^{10}$    & $10$ & $100$	& $ 0.1$	& $46$	& $10^{27}$	& $10$	& $1595$		& $0.5$	& $5.8\,10^{32}$ & $6.4\,10^{34}$	& $6.1\,10^{34}$	& $11.4$	\\ 
\hline\end{tabular}
\end{table*}

\begin{figure}
	\includegraphics[width=\columnwidth]{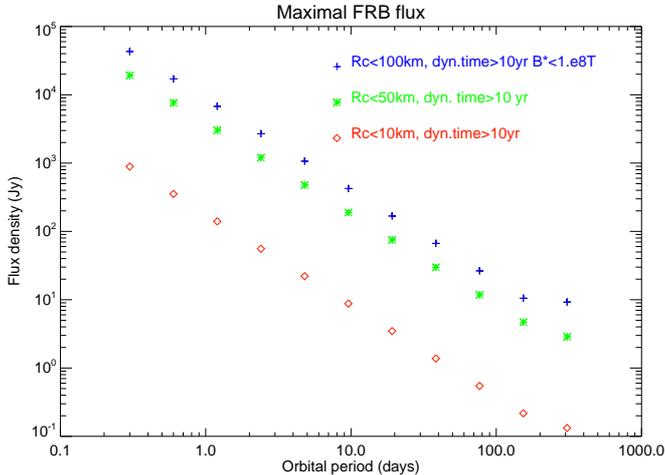}
	\caption{ Maximum observable FRB flux density vs orbital period from the parameter study described in Table \ref{table_parametrique_2}. Each curve corresponds to a different constraints on pulsar spin down age $\tau_\mathrm{sd}$, surface magnetic field $B_*$ and companion radius $R_\mathrm{c}$ (see legend).}
	\label{smax_vs_P}
\end{figure}

{What are the observational constraints 
{that we can derive} ? Of course, observations cannot constrain the various pulsar characteristics intervening in the present parametric study. Nevertheless, in the perspective of the discovery of a periodicity with repeating FRBs, such as with FR180916.J0158+65 \citep{CHIME_2019} and possibly FRB 121102 \citep{Rajwade_2020periodicFRB121102}, it could be interesting to establish a flux/periodicity relation. Figure \ref{smax_vs_P} shows the higher fluxes $S$ as a function of the orbital periods $T_\mathrm{orb}$ from the parameters displayed in Table \ref{table_parametrique_2}. 
Also, the spin-down age 
$\tau_\mathrm{sd}=P_*/2 \dot P_*$ is the characteristic time of evolution of the pulsar period $P_*$. For the non-evaporating asteroids considered in our study, the repeating FRBs can be observed over the time $\tau_\mathrm{sd}$ with constant ranges of fluxes and frequencies. Over a longer duration $t \gg \tau_\mathrm{sd}$, the flux $S$ may fall below the sensitivity threshold of the observations.} 
{Therefore, a prediction of the present model is that repeaters should fade-off with time.}

\subsection{Magnetars and small-size companions} \label{parametrique_magnetars}
We explored parameter sets describing small companions orbiting a magnetar. The list of parameters is displayed in Table \ref{table_parametrique_3}.

{Let us mention that the present list of parameters is intended for FRB that should be observable from a 1 Gpc distance. The recently discovered FRB that is associated to a source (probably a magnetar) in our Galaxy \citep{CHIME_2020_FRBgalactique,Bochenek_2020_radio}, will be the object of a further dedicated study. }

\begin{table*}
	\caption{Sets of input parameters for the parametric study for magnetars.  Only the lines that differ from Table \ref{table_parametrique_2} are plotted.}
	\label{table_parametrique_3} 
	\centering 
	\begin{tabular}{|p{4cm}|p{1.7cm}|l|p{1.2cm}|p{1.5cm}|} 
		\hline 
		Input parameters		& Notation 			& Values 						&	Unit		& Number of values	\\
		\hline
		NS magnetic field 		& $B_*$				& $10^{10+n/2}, n \in\{0,4\}$ 		&	T	 	& 		5		\\		
		NS rotation period		& $P_*$				& $0.1, 0.32, 1, 3.2, 10$ 			&	s		& 		5		\\
		NS temperature		& $T_*$				& $6.\,10^5, 1.9\,10^6, 6\,10^6$	&	K		& 		3		\\
		Companion radius		& $R_\mathrm{c}$				& 1, 2.2, 4.6, 10, 22, 46, 100,  316, 1000, 3162, 10000			&	km	& 	11		\\
		Companion conductivity	& $\sigma_\mathrm{c}$			& $10^{-2}, 10, 10^{4}, 10^7 $ 		&	mho	 	& 		4		\\
		Power input		& $(1-f) gL_\mathrm{sd}$	& $10^{27+n/2}, n \in\{0,10\}$		&	W	 	& 	 	11		\\
		\hline
	\end{tabular}
\end{table*}

For each value of the radio-efficiency $\epsilon$, this corresponds to 15,681,600 parameter sets. 
With $\epsilon=10^{-2}$, we got 20,044 solutions fitting the minimal requirements defined in section \ref{sec_parametric}.
All of them require a minimum size $R_\mathrm{c}=46$ km (from our parameter set), and a very short magnetar period $P_* \le 0.32$ s. Therefore the spin-down age $\tau_\mathrm{sd}$ does not exceed 10 or 20 years. An example is given in case 16 of Table \ref{table_jeux_de_parametres_Rc_petit}.
Therefore, it might be possible to trigger occasional FRBs with very young magnetars, but magnetars at 1 Gpc distance are not the best candidates for repeating FRBs associated with an asteroid belt.

\subsection{Compatibility with the high Faraday rotation measure of FRB 121102} \label{sec_faraday}

\begin{figure}
	\includegraphics[width=\columnwidth]{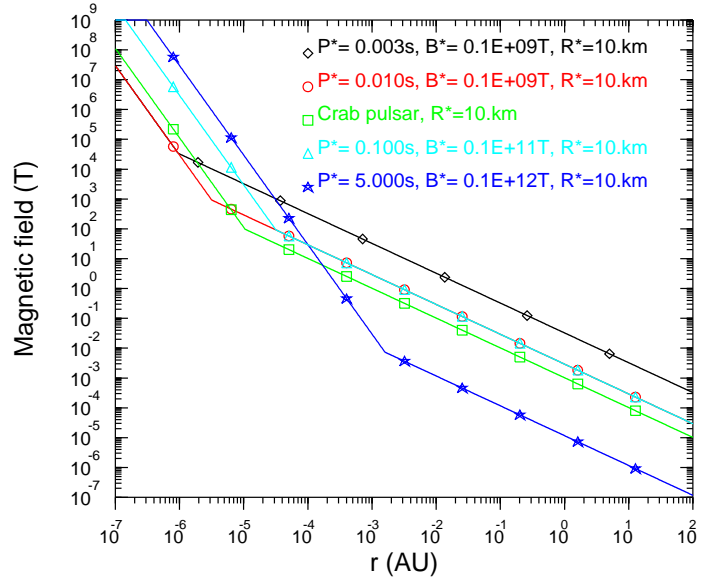}
	\caption{ Magnetic field as a function of the distance from the neutron star from Eq. (\ref{eq_B_dipole}). We can see that beyond $10^{-2}$ AU, the magnetic field is larger in the vicinity of a young pulsar (for instance the Crab pulsar) than for a typical magnetar (blue curve). }
	\label{fig_B_vs_distance}
\end{figure}

\citet{Michilli_2018} have reported observations of 16 bursts associated with FRB 121102, at frequencies 4.1-4.9 GHz with the Arecibo radio-telescope All of them are fully linearly polarized. The polarization angles $PA$ have a dependency in $f^{-2}$ where $f$ is the wave frequency. This is interpreted as Faraday effect. According to the theory generally used by radio astronomers,  when a linearly polarized wave propagates through a magnetized plasma of ions and electrons, its polarization angles $ PA$ in the source reference frame varies as $PA= PA_\infty +\theta = PA_\infty +RM c^2/ f^2$ where $PA_\infty$ is a reference angle at infinite frequency. The $RM$ factor is called the rotation measure and, given in rad.m$^{-2}$,
\begin{equation} \label{eq:def:RM}
RM = 0.81 \int_{D}^{0} B_\parallel(l) n_\mathrm{e} (l) \mathrm{d}l,
\end{equation}
where $B_\parallel$ is the magnetic field  in $\mu$G projected along the  line of sight, $l$ is the distance in parsecs, and $n_\mathrm{e}$ is the electron number density in cm$^{-3}$.

\citet{Michilli_2018}  reported very large values $RM_\mathrm{obs} = (+1.027 \pm 0.001) \times 10^5$ rad m$^{-2}$. With the cosmological expansion redshift $RM_\mathrm{source}= RM_\mathrm{obs} (1+z)^2$ and $z = 0.193$, we have $RM_\mathrm{source} = 1.46 \times 10^5$ rad m$^{-2}$.
The $RM_\mathrm{obs}$ are so large that they could not be detected at lower frequencies, 1.1-2.4 GHz, because of depolarization in the relatively coarse bandwidths of the detectors.  
Measurements performed later at the Green Bank Telescope at 4-8 GHz provided similar values: $RM_\mathrm{source} =  1.33 \times 10^5$ rad m$^{-2}$ \citep{Gajjar_2018}. 

Let us first mention that a pair-plasma does not produce any rotation measure. Therefore, these $RM$ are produced in an ion-electron plasma, supposed to be present well beyond the distances $r$ of the pulsar companions of our model.  
\citet{Michilli_2018}  propose that the rotation measure would come from a 1 pc HII region of density $n_\mathrm{e} \sim 10^{2}$ cm$^{-3}$. This would correspond to an average magnetic field $B_{\parallel} \sim$ 1 mG. But for comparison, average magnetic fields in similar regions in our Galaxy correspond typically to 0.005 mG. Therefore, it is suggested that the source is in the vicinity of a neutron star and, more plausibly, of a magnetar or a black hole. 

{Figure \ref{fig_B_vs_distance} is a plot of the magnetic field as a function of the distance from the neutron star. It is approximated in the same way as in the rest of the paper : a dipole field up to the light-cylinder, then, a magnetic field dominated by the toroidal component $B_\phi$ that decreases as $r^{-1}$ (see Eq. \ref{eq_B_dipole}). It is plotted for a highly magnetized magnetar ($P_*=5$ s, $B_*=10^{12}$ T), for a Crab-like pulsar, for the very young pulsars that fit the parametric study of section \ref{parametrique_pulsars}, and for a fast and short lived magnetar ($P_*=0.1$ s, $B_*=10^{11}$) that could provide FRBs with $R_\mathrm{c}=50$ km companions. We can see that at close distances from the neutron star, the highly magnetized magnetar as well as the fast magnetar have the largest magnetic fields. But since their light-cylinder is farther away than for fast young pulsars, the magnetar's field beyond a distance $r=0.01$ AU is lower than that caused by fast young pulsars. }

{Therefore, we can conclude that the high interstellar magnetic field in the 1 pc range of distances, that can be the cause of the large rotation measure associated with FRB 121102 is likely to be produced by the young pulsars  that fit the quantitative model presented in section \ref{parametrique_pulsars}.}

\section{Effects of the wind fluctuations on burst duration and multiplicity} \label{sec_duration_multiplicity}

The magnetic field is not constant in the pulsar wind, and its oscillations can explain why radio emission associated with a single asteroid cannot be seen at every orbital period $T_\mathrm{orb}$ or, on the contrary, could be seen several times.
The direction of the wind magnetic field oscillates at the spin period $P_*$ of the pulsar which induces a slight  variation of the axis of the emission cone due to relativistic aberration (see MZ14). In the reference frame of the source, the important slow oscillation (period $P_*$) of the magnetic field can induce stronger effects on the plasma instability, and therefore, on the direction of the emitted waves. In the observer's frame, this will consequently change the direction of emission \textit{within} the cone of relativistic aberration. 
These small changes of direction combined with the long distance between the source and us will affect our possibility of observing the signal. According to the phase of the pulsar rotation, we may, or may not actually observe the source signal. It is also possible, for a fast rotating pulsar, that we observe separate peaks of emission, with a time interval $P_* \sim 1$ ms or more. Actually, bursts with such multiple peaks of emission  (typically 2 or 3) have been observed with FRB 181222, FRB 181226, and others \citep{CHIME_2019} and FRB 121102 \citep{spitler_repeating_2016}. 

The image of a purely stationary Alfv\'en wing described up to now in our model is of course a simplification of reality. Besides the variations due to the spinning of the pulsar, the wind is anyway unlikely to be purely stationary: the energy spectrum of its particles varies, as well as the modulus and orientation of the embedded magnetic field. 
More importantly, since the flow in the equatorial plane of the neutron star is expected to induce magnetic reconnection at a distance of a few AU, we expect that turbulence has already started to develop at 0.1 AU (and probably even at 0.01 AU), so that fluctuations of the wind direction itself are possible. {As a consequence, the wind direction does not coincide with the radial direction, but makes a small and fluctuating angle with it.} These angular fluctuations must result in a displacement of the source region as well as of the direction of the emission beam. This wandering of the beam is bound to affect not only burst durations but also their multiplicities.

\subsection{Duration}\label{burst_duration}

Two characteristics must be discussed regarding the duration of the bursts : the narrowness and the relatively small dispersion of their durations. For the first 17 events associated with FRB121102 in the FRB Catalog \citep{Petroff16}\footnote{http://www.frbcat.org}, the average duration is 5.1 ms and the standard deviation is 1.8 ms\footnote{We can notice that for all FRB found in the catalog, where FRB121102 counts for 1 event, the average width, 5.4 ms, is similar to that of FRB121102, and the standard deviation is  6.3 ms}. 

Considering a circular Keplerian orbit of the NS companion, the time interval $\tau$  during which radio waves are seen is the sum of a time $\tau_{\rm beam}$ associated with the angular size of the beam $\alpha_{\rm beam}$ (width of the dark green area inside the green cone in fig. \ref{fig_source_size}), and a time $\tau_{\rm source}$ associated with the size of the source $\alpha_{\rm source} $ (size of the yellow region in fig. \ref{fig_source_size}). In MZ14, the term $\tau_{\rm source}$ was omitted, leading to underestimate $\tau$. We take it into account here.

Assuming for simplicity that the emission beam is a cone of aperture angle $\alpha_{\rm beam}$ then the corresponding solid angle is $\Omega_{\rm beam} = \pi\alpha_{\rm beam}^2/4$. Using Eq. \eqref{eq:omegabeam}, we obtain, 
\begin{equation}
\alpha_{\rm beam} = \frac{1}{\gamma} \left(\frac{\Omega_{\rm A}}{\pi}\right)^{1/2} = 2\times 10^{-7} {\rm rad} \left(\frac{10^6}{\gamma}\right)\left(\frac{\Omega_{\rm A}}{0.1 {\rm sr}}\right)^{1/2},
\end{equation}
where $\Omega_{\rm A}$ is the intrinsic solid angle of emission defined in Sect. \ref{section_solid_angle}.

Similarly, the size of the source covers an angle $\alpha_{\rm source} \sim R_{\rm s}/r$ where $R_S$ is the size of source and $r$ the distance of the companion to the pulsar. Assuming a $1.4M_\odot$ neutron star, this leads to 
\begin{equation}
\alpha_{\rm source} = 3\times 10^{-7}\, {\rm rad} \left(\frac{R_{\rm s}}{10\,{\rm km}}\right) \left(\frac{T_{\rm orb}}{0.1\, {\rm yr}}\right)^{2/3}.
\end{equation}
The angle $\alpha_{\rm source}$ is an upper limit, since the source can be further down the wake. 

Assuming the line of sight to lie in the orbital plane, then the transit time is equal to the time necessary for the line of sight to cross the angle of emission at a rate $n_{\rm orb} = 2\pi/T_{\rm orb}$ such that $n_{\rm orb}\tau_{\rm (beam/source)} = \alpha_{\rm (beam/source)}$, giving 
\begin{eqnarray}
\tau \hspace{6mm} & = & \tau_{\rm beam}+\tau_{\rm source} \nonumber \\
\tau_{\rm beam} & = & 0.09 \mathrm{s} \left(\frac{T_{\rm orb}}{0.1\, {\rm yr}}\right) \left(\frac{10^6}{\gamma}\right)\left(\frac{\Omega_{\rm A}}{0.1 {\rm sr}}\right)^{1/2}, \nonumber \\
\tau_{\rm source} & = & 0.13 \mathrm{s} \left(\frac{R_{\rm s}}{10\,{\rm km}}\right) \left(\frac{T_{\rm orb}}{0.1\, {\rm yr}}\right)^{4/3},
\end{eqnarray}
which should be seen as maximum transit durations since, in this simplified geometry, the line of sight crosses the widest section of the beam. Nonetheless, one sees that $\tau \sim 0.2$ s largely exceeds the average duration of 5 ms that has been recorded for FRB121102.

This discrepancy can be resolved if one considers the {angular} wandering of the source due to turbulence and intrinsic wind oscillations. Let $\dot\omega$ be the characteristic angular velocity of the beam as seen from the pulsar in the co-rotating frame. One now has $(n_{\rm orb} + \dot \omega)\tau = \alpha_{\rm beam} +\alpha_{\rm source}$, where $\tau$ is the observed value $\tau \sim 5$ ms. This gives $\dot \omega \simeq 10^{-4} \mathrm{rad/s}$ (or 0.7 days in terms of period) for $T_{\rm orb} =0.1\, {\rm yr}, R_{\rm s} =10\,{\rm km}, \Omega_{\rm A} =0.1 \,{\rm sr}, \gamma =10^6$. At a distance $\sim r$ this corresponds to a velocity of the source of $v_s \simeq  0.01c$, which is much smaller than the radial velocity of the wind, $c$, and therefore seems plausible. Because one does not generally expect that a burst corresponds to crossing the full width of the beam, this source velocity is to be seen as an upper limit.

\subsection{Bursts multiplicity and rate} \label{sec_discussion_multiplicity}

As underlined in \citet{Connor_2016}, a non-Poissonian repetition rate of bursts, as is observed, can have important consequences on the physical modeling of the sources. Actually, it is remarkable that many bursts come in groups gathered in a time interval of a few minutes. \citet{Zhang_2018}  used neural network techniques for the detection of 93 pulses from FRB 121102 in 5 hours, 45 of them being detected in the first 30 minutes. Some of them repeated within a few seconds. Many were short (typically 1 to 2 ms duration) and had low amplitudes ($S$ frequently below 0.1 Jy).

We are therefore considering in this section that the bunching of pulses over a wandering timescale $\tau_{\rm w} \sim 1$h is due to the turbulent motion of the beam of a single asteroid crossing several times the line of sight of the observer. During that time the asteroid moves on its orbit by an angle $\alpha_{\rm w} = n_{\rm orb} \tau_{\rm w} \gg \alpha$ (see Sect. \ref{burst_duration}). In particular, 
\begin{equation}
\label{eq:alphaw}
\alpha_{\rm w} = 7. 10^{-3} {\rm rad} \left(\frac{ \tau_{\rm w}}{1\, {\rm h}}\right) \left(\frac{T_{\rm orb}}{0.1\, {\rm yr}}\right)^{-1},
\end{equation}
which implies that the beam is only wandering over a small region. For the fiduciary values used in this formula, this corresponds to $\alpha_{\rm w}= 0.4^\circ$. The fact that several bursts are visible also suggests that the beam covers the entire region during that time interval, possibly several times. This leads to associate an effective solid angle to the emission of an asteroid, that is the solid angle within which emission can be detected with a very high probability provided that one observes for a duration $\sim \tau_{\rm w}$. Further considering that this effective solid angle is dominated by the effect of wandering, one may define 
\begin{equation}
\Omega_{\rm w} = \frac{\pi}{4}\alpha_{\rm w}^2,
\end{equation}
where we have assumed a conical shape which is expected, for instance, in case of isotropic turbulence.
We also note that, in order to see two bursts separated by $\tau_{\rm w}$, the source must travel at a speed $v_s > r n_{\rm orb} = 2\times 10^{-4}c (T_{\rm orb}/0.1{\rm yr})^{-1/3}$ compatible with the findings of Sect. \ref{burst_duration}.

Thus, we propose that groups of bursts observed within a time interval of the order of $\tau_{\rm w}$ result from one asteroid--pulsar interaction.
Therefore, when estimating bursts rates, it is important to distinguish the number of bursts from the number of burst groups. According to our model, only the latter corresponds to the number of neutron star-asteroid-Earth alignments. We emphasize that the meandering of the direction of emission is an unpredictable function of time connected with the turbulent flow which suggests that one will not observe precise periodic repetitions at each asteroid alignment (this effect comes in addition to the gravitational interactions mentioned in section \ref{sec_clusters_belts}).
We also note that the bunching mechanism we propose here does not predict a particular pattern for the properties of each bursts, which may vary randomly from one to another. This is distinct from an eruption-like phenomenon where an important release of energy is followed by aftershocks of smaller intensity, such as predicted by models of self-organized criticality \citep{aschwanden_25_2016}.

\section{Discussion}

\subsection{What is new since the 2014 model with neutron stars and companions} \label{sec_quoideneuf}
MZ14, revised with MH20, showed that planets orbiting a pulsar, in interaction with its wind could cause FRBs. Their model predicts that FRBs should repeat periodically with the same period as the orbital period of the planet. An example of possible parameters given in MZ14 was a planet with a size $\sim 10^4$ km at 0.1 AU from a highly magnetized millisecond neutron star.
Later, \citet{Chatterjee_2017} published the discovery of the repeating FRB121102. The cosmological distance of the source (1.7 Gpc) was confirmed. But the repeater FRB121102, as well as the others discovered since then, are essentially not periodic.

In the present paper, we consider that irregular repeating FRBs can be triggered by belts or swarms of small bodies orbiting a pulsar. 
After adding some thermal considerations that were not included in MZ14, we conducted several parametric studies that showed that repeating FRBs as well as non-repeating ones can be generated by small bodies in the vicinity of a very young pulsar, and less likely, 
a magnetar. Some parameters sets even show that 10 km bodies could cause FRBs seen at a distance of 1 Gpc with a flux of tens of Janskys.

In MZ14, it was supposed that the radio waves could be emitted by the cyclotron maser instability. But the frequency of CMI waves is slightly above the electron gyrofrequency that, with the retained parameters, is not compatible with observed radio frequencies. 
Therefore, another emission mechanism must be at play.
{Determining its nature is left for a further study.}

Whatever the radio emission generation process, our parametric study is based on the hypothesis that in the source reference frame, the 
the emission is modestly beamed within 
a solid angle $\Omega_\mathrm{A} \sim 0.1$ to 1 sr. 
It will be possible to refine our parametric study when we have better constrains
 on $\Omega_\mathrm{A}$.

\subsection{Comparisons with other models of bodies interacting with a neutron star}

Many kinds of interactions have been studied between a neutron stars and celestial bodies orbiting it. 
Many of them involve interactions between the neutron star and its companion at a distance large enough for dissipation of a significant part of the pulsar wind magnetic energy. The pulsar wind energy is dominated by the kinetic energy of its particles. Moreover, the companion has an atmosphere, or a wind of its own if it is a star, and the companion-wind interface includes shock-waves. 

In the present model, the distance between the companion is much smaller, the wind energy is dominated by magnetic energy. The companion is asteroid-like, it has no atmosphere, and whatever the Mach number of the pulsar wind, there is no shock-wave in front of it. The wind-companion interface takes the form of an Alfv\'en wing.   

The present model is not the only one involving asteroids in the close vicinity of a neutron star. \citet{Dai16} consider the free fall of an asteroid belt onto a highly magnetized pulsar, compatible with the repetition rate of FRB121102. The asteroid belt is supposed to be captured by the neutron star from another star (see \citet{Bagchi_2017} for capture scenarios). The main energy source is the gravitational  energy release related to tidal effects when the asteroid passes through the pulsar breakup radius.
 After \citet{Colgate_1981}, and with their fiduciary values, they estimate this power to $\dot E_\mathrm{G} \sim 1.2 \, 10^{34}$ W. Even if this power is radiated isotropically from a source at a cosmological distance, this is enough to explain the observed FRB flux densities. \citet{Dai16} developed a model to explain how a large fraction of this energy is radiated in the form of radio waves. 
 As in MZ14, they invoke the unipolar inductor model. Their electric field has the form $\vec E_2 = - \vec v_\mathrm{ff} \times \vec B$ {where $\vec v_\mathrm{ff}$ is the free-fall asteroid velocity} (we note it $E_2$ as in \citet{Dai16}.) 
Then, the authors argue that this electric field can accelerate electrons up to a Lorentz factor $\gamma \sim 100$, with a density computed in a way similar to the Goldreich-Julian density in a pulsar magnetosphere.  
These accelerated electrons would cause coherent curvature radiation at frequencies of the order of $\sim 1$ GHz. Their estimate of the power associated with these radio waves is $\dot E_\mathrm{radio} \sim 2.6 \, \times 10^{33}$ W $\sim 0.2\, \dot E_\mathrm{G}$, under their assumption (simple unipolar inductor) that partially neglects the screening of the electric field by the plasma (contrarily to the AW theory) and therefore overestimates the emitted power.

{Conceptually, the \citet{Dai16} model is more similar to ours than models related to shocks. In terms of distances  to the neutron star, our model places the obstacle (asteroid, star) in between these two.}

\subsection{How many asteroids ?} \label{sec_how_many_asteroids}
 
Let us consider a very simple asteroid belt where all the asteroids have a separation $r$ with the neutron star corresponding to an orbital period $T_{orb}$, and their orbital inclination angles are less than $\alpha=\Delta z/r$ where $\Delta z$ is the extension in the direction perpendicular to the equatorial plane of the star.

Let $N_{\rm v}$ be the number of ``visible'' asteroids whose beams cross the observer's line of sight at some point of their orbit.
As explained in Sect. \ref{sec_discussion_multiplicity}, we may consider that due to the {angular} wandering of the beam, an asteroid is responsible for all the bursts occurring with a time window of $\tau_{\rm w} \sim 1$h. Assuming that the beam covers a conical area of aperture angle $\alpha_{\rm w}$ defined in Eq. \eqref{eq:alphaw}, the visible asteroids are those which lie within $\pm \alpha_{\rm w}/2$ of the line of sight of the observer at inferior conjunction, the vertex of angle being the pulsar. The total number $N$ of asteroids in the belt is then related to the number of visible asteroids by $N = N_{\rm v} \alpha/\alpha_{\rm w}$, assuming a uniform distribution.
By definition, the number of groups of bursts per orbital period is $N_{\rm v}$. It follows that the number of groups during a time $\Delta t$ is given by $N_{g} = N_{\rm v}  {\Delta t}/T_{orb}$. Defining the rate $n_{\rm g} = N_{\rm g}/\Delta t$, we obtain $N = n_{\rm g} T_{\rm orb} \alpha / \alpha_{\rm w}$. 

About 9 groups of pulses separated by more than one hour are listed in FRBcat for FRB121102. Some FRB repeaters have been observed for tens of hours without any burst, the record being 300 hours of silence with  FRB 180814.J0422+73 \citep{Oostrum_2020}. Considering an interval of 100 hours between bunches of bursts, the event rate per year would be about 88. Let us adopt the figure of 100 events per year, giving 
\begin{equation} \label{eq_N_asteroides}
N = 1.4\times 10^2 \left(\frac{n_{\rm g}}{100\, {\rm yr^{-1}}}\right) \left(\frac{\alpha}{0.1{\rm rad}}\right)\left(\frac{\tau_{\rm w}}{1\, {\rm h}}\right)^{-1} \left(\frac{T_{\rm orb}}{0.1 {\rm yr}}\right)^2.
\end{equation}
Therefore, for the fiduciary values in Eq. (\ref{eq_N_asteroides}), only 140 asteroids are required to explain a FRB rate similar to those of the repeater FRB121102.
For comparison, in the solar system, more than $10^6$ asteroids larger than 1 km are known, of which about $10^4$ have a size above 10 km.

\section{Conclusion and perspectives}

{Here is a summary of the characteristics of FRBs explained by the present model.
Their observed flux (including the 30 Jy Lorimer burst-like) can be explained by a moderately energetic system because of a strong relativistic aberration associated with a highly relativistic pulsar wind interacting with metal-rich asteroids. A high pulsar wind Lorentz factor ($\sim 10^6$) is required in the vicinity of the asteroids. The FRB duration (about 5 ms)  is associated with the asteroid orbital motion, and the periodic variations of the wind caused by the pulsar rotation, and to a lesser extend  with pulsar wind turbulence. The model is compatible with the observations of FRB repeaters. But the repetition rate could be very low, and FRB considered up to now as non-repeating could be repeaters with a low repetition rate. For repeaters, the bunched repetition rate is a consequence of a moderate pulsar wind turbulence. All the pulses associated with the same neutron star should have the same DM (see appendix \ref{section_fluctuation_DM}). The model is compatible with the high Faraday rotation measure observed with FRB 121102. 
Of course, the present model involves the presence of asteroids orbiting a young neutron star. It is shown with the example of FRB 121102 that less than 200  asteroids with a size of a few tens of km is required. This is much less than the number of asteroids of this kind orbiting in the solar system. }

 Nevertheless, we do not pretend to explain all the phenomenology of FRBs, and it is possible that various FRBs are associated with different families of astrophysical sources. 
 
{We can also mention perspectives of technical improvements of pulsar--asteroids model of FRB that deserve further work.}

{As mentioned in section \ref{sec_quoideneuf}, the radio emission process is left for a further study.}

When analyzing thermal constraints, we have treated separately the Alfv\'en wing and the heating of the companion by the Poynting flux of the pulsar wave.  For the latter, we have used the Mie theory of diffusion which takes into account  the variability of the electromagnetic environment of the companion, but neglects the role of the surrounding plasma. On another hand, the Alfv\'en wing theory, in its present state, takes the plasma into account, but neglects the variability of the electromagnetic environment. A unified theory of an Alfv\'en wing in a varying plasma would allow a better description of interaction of the companion with its environment.

Another important (and often asked) question related to the present model is the comparison of global FRB detection statistics with the population of young pulsars with asteroids or planets orbiting them in galaxies within a 1 or 2 Gpc range from Earth. This will be the subject of a forthcoming study.

{Since our parametric study tends to favor 10 to 1000 years old pulsars, the FRB source should be surrounded by a young expanding supernova remnant (SNR), and a pulsar wind nebula driving a shock into the ejecta \citep{Gaensler_2006}. The influence of the SNR plasma onto the dispersion measure of the radio waves will be as well the topic of a further study. This question requires a model of the young SNR, and it is beyond the scope of the present paper. Anyway, we cannot exclude that this point bring new constraints on the present model, and in particular on the age of the pulsars causing the FRBs or on the mass of its progenitor.
}

We can end this conclusion with a vision. 
After upgrading the MZ14 FRB model (corrected by MH20) with the ideas of a cloud or belt of asteroids orbiting the pulsar, and the presence of turbulence in the pulsar wind, we obtain a model that explains observed properties of FRB repeaters and non-repeaters altogether in a unified frame. In that frame, the radio Universe contains myriads of pebbles, circling around young pulsars in billions of galaxies, each have attached to them a highly collimated beam of radio radiation, meandering inside a larger cone (of perhaps a fraction of a degree), that sweeps through space, like multitudes of small erratic cosmic lighthouses, detectable at gigaparsec distances. Each time one such beam crosses the Earth for a few milliseconds and illuminates radio-telescopes, a FRB is detected, revealing an ubiquitous but otherwise invisible reality.

\section*{Acknowledgment}
G. Voisin acknowledges support of the European Research Council, under the European Union’s Horizon 2020 research and innovation programme (grant agreement No. 715051, Spiders). 

\appendix
\section{Another way of computing the radio flux} \label{section_AW_bis}
As shown in \citet{Mottez_2011_AWW}, the optimal power 
of the Alfv\'en wing $\dot E_\mathrm{A}$ is the fraction of the pulsar Poynting flux intercepted by the companion section. 
Considering that, at the companion separation $r$, most of the pulsar spin-down power $L_\mathrm{sd}$ 
is in the form of the Poynting flux,  $L_\mathrm{sd}=I \Omega_* \dot \Omega_*$ is a correct estimate of the Poynting flux. It can be derived from our parameter set. The fiduciary value of inertial momentum is  $I=10^{38}$ kg.m$^2$. Our parameters do not include explicitly $\dot \Omega_*$. Fortunately, we can use the approximate expression
\begin{equation}
{L_\mathrm{sd}} = \frac{\Omega_*^4 B_*^2 R_*^6}{c^3} (1 + \sin^2 i)
\end{equation}
where $i$ is the  magnetic inclination angle \citep{Spitkovsky_2006}. We choose the conservative value $i=0^\circ$. 
With normalized values, 
\begin{equation}\label{Erp}
\left( \frac{L_\mathrm{sd}}{\mbox{W}}\right)= 5.8 \times 10^{26} \left( \frac{R_*}{10\mbox{ km}}\right)^6  \left( \frac{B_*}{10^5 \mbox{T}}\right)^2  \left( \frac{10\mbox{ ms}}{P_*}\right)^4
\end{equation}
Following Eq. (\ref{puissance_radio}), the radio power is $\dot E_\mathrm{R}'=\epsilon \dot E_\mathrm{A}'$, where the prime denotes the present alternative computation. 
\begin{equation}\label{radio_power_spin_down}
\dot E_\mathrm{R}'=\epsilon  L_\mathrm{sd} \frac{\pi R_\mathrm{c}^2}{4 \pi r^2}.
\end{equation} 
The equivalent isotropic luminosity is 
\begin{equation}
\dot E_\mathrm{iso}'=\dot E_\mathrm{R}' \frac{4 \pi}{\Omega_\mathrm{beam}}=  \epsilon L_\mathrm{sd} \frac{R_\mathrm{c}^2}{r^2} \gamma^2 \frac{4 \pi}{\Omega_\mathrm{A}},
\end{equation}
where again, the prime means ``alternative to Eq. (\ref{EisoS})''. In terms of reduced units,
\begin{equation}\label{EisoA}
\left( \frac{\dot E_\mathrm{iso}'}{\mbox{W}}\right)= 5.52 \times 10^2 \left( \frac{L_\mathrm{sd}}{\mbox{W}}\right)  \left( \frac{R_\mathrm{c}}{10^4\mbox{km}} \right)^2 \left( \frac{AU}{\mbox{r}}\right)^2 \left( \frac{\gamma}{10^5}\right)^2.
\end{equation}
Let us notice that, owing to the relativistic beaming factor $\gamma^2$, this power can exceed the total pulsar spin-down power. This is possible because the radio beam covers a very small solid angle, at odds with the spin-down power that is a source of Poynting flux in almost any direction. 
Equation (\ref{E_iso_prime}) is a simple combination of Eqs. (\ref{Erp}) and (\ref{EisoA}). 

\section{Asteroid heating} \label{section_heat_sources}
\subsection{Heating by thermal radiation from the neutron star} \label{section_thermal}
For heating by the star thermal  radiation, 
\begin{equation} \label{L_vers_E}
\dot E_\mathrm{T}= \frac{L_\mathrm{T}}{4}  \left(\frac{R_\mathrm{c}}{r}\right)^2  \mbox{ and } L_\mathrm{T}=4 \pi \sigma_\mathrm{S} R_*^2  T_*^4
\end{equation}
where $L_\mathrm{T}$ is the thermal luminosity of the pulsar, $R_*$ and $T_*$ are the neutron star radius and temperature, $\sigma_\mathrm{S}$ is 
the Stefan-Boltzmann constant and in international system units, $4 \pi \sigma_\mathrm{S}= 7 \times 10^{-7}$ Wm$^{-2}$K$^{-4}$. For $T_*=10^6$ K, and $T_*=3 \times 10^5$, the luminosities are respectively $7 \times 10^{25}$ W and $10^{24}$ W.
This is consistent with the thermal radiation of Vela observed with Chandra, $L_\mathrm{T} = 8 \times 10^{25}$ W \citep{Zavlin_2009}.

\subsection{Heating by the pulsar wave Poynting flux} \label{section_poynting}
Heating by the Poynting flux has been investigated in \citet{Kotera_2016}. Following their method,  
\begin{equation} \label{eq_E_P_base}
\dot E_\mathrm{P}= \left(\frac{f L_\mathrm{sd}}{f_\mathrm{p}}\right)\left(\frac{ Q_\mathrm{abs}}{4}\right)\left(\frac{R_\mathrm{c}}{r}\right)^2
\end{equation}
where  $L_\mathrm{sd}$ 
is the loss of pulsar rotational energy.
The product $f L_\mathrm{sd}$ is the part of the rotational energy loss that is taken by the pulsar wave.
The dimensionless factor $f_\mathrm{p}$ is the fraction of the sky into which the pulsar wind is emitted. 
The absorption rate $Q_\mathrm{abs}$ of the Poynting flux by the companion is derived in \citet{Kotera_2016} by application of the Mie theory with the Damie code based on \citet{Lentz_1976}. For a metal rich body whose size is less than $10^{-2} R_\mathrm{LC}$, $Q_\mathrm{abs} < 10^{-6}$ (see their Fig. 1). We note
$Q_\mathrm{max}=10^{-6}$ this upper value.

The value $Q_\mathrm{{max}}=10^{-6}$ may seem surprisingly small. Indeed, for a large planet, we would have $Q_\mathrm{abs} \sim 1$. To understand well what happens, we can make a parallel with the propagation of electromagnetic waves in a dusty interstellar cloud. The size of the dust grains is generally $\sim$ 1 $\mu$m. Visible light, with smaller wavelengths ($\sim 500 $ nm) is scattered by these grains. But infrared light, with a wavelength larger that the dust grains is, in accordance with the Mie theory, not scattered, and the dusty cloud is transparent to infrared light. 
With a rotating pulsar, the wavelength of the ``vacuum wave'' (Parker spiral) is $\lambda \sim 2 c/\Omega_*$ equal to twice the light cylinder radius $r_\mathrm{LC}=c/\Omega_*$. 
Dwarf planets are comparable in size with the light cylinder of a millisecond pulsar and much smaller than that of a 1s pulsar, therefore smaller bodies always have a factor Q smaller or much smaller than 1. Asteroids up to 100 km are smaller than the light cylinder, therefore they don't scatter, neither absorb, the energy carried by the pulsar ``vacuum'' wave. This is what is expressed with $Q_\mathrm{{max}}=10^{-6}$. 

{If the companion size is similar to the pulsar wavelength, that is twice the light-cylinder distance $2 r_\mathrm{LC}$, then heating by the Poynting flux cannot be neglected. In that case, we can take $Q_\mathrm{{abs}} =1$. Therefore, for medium size bodies for which gravitation could not retain an evaporated atmosphere, it is important to check if $R_\mathrm{c} < 2 r_\mathrm{LC}$. } 

\subsection{Heating by the pulsar non-thermal radiation} \label{section_non_thermal}

For heating by the star non-thermal radiation, 
\begin{equation} \label{NT_vers_E}
\dot E_\mathrm{NT}= g_\mathrm{NT} \frac{L_\mathrm{NT}}{4}  \left(\frac{R_\mathrm{c}}{r}\right)^2  
\end{equation}
where $g_\mathrm{NT}$ is a geometrical factor induced by the anisotropy of non-thermal radiation. In regions above non-thermal radiation sources, $g_\mathrm{NT}>1$, but in many other places, $g_\mathrm{NT}<1$. 
The total luminosity associated with non-thermal radiation $L_\mathrm{NT}$ depends largely on the physics of the magnetosphere, and it is simpler to use measured fluxes than those predicted by models and simulations.

With low-energy gamma-ray-silent pulsars, most of the energy is radiated in X-rays. Typical X-ray luminosities are in the range $L_\mathrm{X} \sim 10^{25} - 10^{29}$ W, corresponding to a proportion $\eta_\mathrm{X} = 10^{-2}-10^{-3}$ of the spin-down luminosity $L_\mathrm{sd}$ \citep{Becker_2009}.

Gamma-ray pulsars are more energetic. It is reasonable to assume gamma-ray pulsar luminosity $L_{\gamma} \sim 10^{26} - 10^{29}$~W \citep{Zavlin_2009,fermi_2_catalog}. The maximum of these values can be reached with young millisecond pulsars. For instance, the the Crab pulsar (PSR B0531+21), with a period $P_*=33$~ms and $3.8\times10^8$~T
has the second largest known spin-down power $4.6\times10^{31}$~W
and its X-ray luminosity is $L_\mathrm{X} = 10^{29}$~W \citep{Becker_2009}. Its gamma-ray luminosity, estimated with Fermi is $L_\gamma =6.25 \times 10^{28}$~W above 100 MeV \citep{Abdo2010_Crab}. Vela ($P_*=89$~ms, $B_*=3.2\times10^8$~T) has a lower luminosity: $L_\mathrm{X}=6.3\times10^{25}$~W in X-rays (including the already mentioned contribution of thermal radiation) and $L_\gamma=8.2 \times 10^{27}$ ~W in gamma-rays \citep{Abdo_2010_Vela}. 

Concerning the geometrical factor $g_\mathrm{NT}$, it is important to notice that most of the pulsar X-ray and gamma-ray radiations are pulsed. 
Because this pulsation is of likely geometrical origin, their emission is not isotropic. {Without going into the diversity of models that have been proposed to explain the high-energy emission of pulsars it is clear that i) the high-energy flux impinging on the companion is not constant and should be modulated by a duty cycle and ii) the companion may be located in a region of the magnetosphere where the high-energy flux is very different from what is observed, either weaker or stronger. It would be at least equal to or larger than the inter-pulse level.}

{If magnetic reconnection takes place in the stripped wind (see \citet{Petri_revue_2016} for a review), presumably in the equatorial region, then a fraction of the Poynting flux $L_\mathrm{P}$ should be converted into high-energy particles, or X-ray and gamma-ray photons (or accelerated leptons, see section \ref{section_wind_particles}), which would also contribute to the irradiation.}
At what distance does it occur ? Some models based on a low value of the wind Lorentz factor ($\gamma =250$ in \citet{Kirk_2002}) evaluate the distance $r_\mathrm{diss}$ 
of the region of conversion at 10 to 100 $r_\mathrm{LC}$ {(we note ``diss'' for dissipation of magnetic energy)}. With pulsars with $P \sim 0.01$ or $0.1$ s, as we will see later, the companions would be exposed to a high level of X and gamma-rays. 
But, as we will also see, the simple consideration of $\gamma$ as low as 250 does not fit our model. More recently, \citet{Cerutti_2017} showed with numerical simulations a scaling law for $r_\mathrm{diss}$, that writes $r_\mathrm{diss} / R_\mathrm{LC}= \pi \gamma \kappa_\mathrm{LC}$ where $\kappa_\mathrm{LC}$ is the plasma multiplicity at the light cylinder.
{(The plasma multiplicity is the number of pairs produced by a single primary particle.)}
With multiplicities of order $10^3-10^4$ \citep{Timokhin_2013} and  $\gamma > 10^4$ \citep{Wilson_1978,Ng_2018}, and the ``worst case'' $P_*=1$ ~ms, we have $r_\mathrm{diss}>3$ AU, that is beyond the expected distance $r$ of the companions causing FRBs, as will be shown in the parametric study (section \ref{sec_parametric}).

Therefore, we can consider that the flux of high energy photons received by the pulsar companions is less than the flux corresponding to an isotropic luminosity $L_\mathrm{NT}$, therefore we can consider that $g_\mathrm{NT} \le 1$.

\subsection{Heating by the pulsar wind particles that hit the companion}  \label{section_wind_particles}
Heating by absorption of the particle flux can be estimated on the basis of the density of electron-positron plasma that is sent away by the pulsar with an energy $\sim \gamma m_\mathrm{c} c^2$.
Let $n$ be the number density of electron-positron pairs, we can write  $n =\kappa \rho_\mathrm{G} /e$ where $\rho_\mathrm{G}$ is the Goldreich-Julian charge density (also called co-rotation charge density), $\kappa$ is the multiplicity of pair-creations, and $e$ is the charge of the electron. We use the approximation $\rho_\mathrm{G} = 2 \epsilon_0 \Omega_* B_*$. Then,
\begin{equation} \label{eq_L_W}
L_\mathrm{W}= \kappa n_\mathrm{G} \gamma m_\mathrm{e} c^2 4 \pi R_*^2 f_\mathrm{W}
\end{equation}
where $f_\mathrm{W} \le 1$ is the fraction of the neutron star surface above which the particles are emitted.  The flux $\dot E_\mathrm{W}$ is deduced from $L_\mathrm{W}$ in the same way as in Eq. (\ref{L_vers_E}),
\begin{equation}
\dot E_\mathrm{W}=g_\mathrm{W} \frac{L_\mathrm{W}}{4}\left(\frac{R_\mathrm{c}}{r}\right)^2
\end{equation}
where $g_\mathrm{W}$ is a geometrical factor depending on the wind anisotropy. 

\subsection{Added powers of the wind particles and of the pulsar non-thermal radiation} \label{section_wind_and_non_thermal}
We can notice in Eq. (\ref{eq_L_W}) that the estimate of $L_\mathrm{W}$ depends on the ad-hoc factors $\kappa$, $\gamma$ and $f$. Their estimates are highly dependent on the various models of pulsar magnetosphere. 
It is also difficult to estimate $ L_\mathrm{NT}$. 
Then, it is convenient to notice that there are essentially two categories of energy fluxes: those that are fully absorbed by the companion (high-energy particles, photons and leptons), and those that may be only partially absorbed (the Poynting flux). Besides, all these fluxes should add up to the total rotational energy loss of the pulsar $L_\mathrm{sd}$. Therefore, the sum of the high-energy contributions may be rewritten as being simply
\begin{equation} \label{eq_sioux_1}
g_\mathrm{W}  L_\mathrm{W} + g_\mathrm{NT} L_\mathrm{NT}= (1-f) g L_\mathrm{sd} 
\end{equation}
where $f$ is the fraction of rotational energy loss into the pulsar wave, already accounted for in Eq. (\ref{eq_E_P_base}). 
This way of dealing with the problem allows an economy of ad-hoc factors. 
The dependency of $g$ is a function of the inclination $i$ of the NS magnetic axis relatively to the companion orbital plane as
a consequence of the effects discussed in section \ref{section_non_thermal}. Following the discussion of the previous sections regarding $g_\mathrm{W}$ and $g_\mathrm{NT}$, we assume generally that  $g$ is less than one. 

\subsection{Heating by the companion Alfv\'en wing} \label{section_Joule_AW}
There is still one source of heat to consider : the Joule dissipation associated with the Alfv\'en wing electric current. 
The total power associated with the Alfv\'en wing  is 
\begin{equation} \label{puissance_AW}
\dot E_\mathrm{A} \sim {I_\mathrm{A}^2}{\mu_0 V_\mathrm{A}} = {I_\mathrm{A}^2}{\mu_0 c}.
\end{equation}
where $V_\mathrm{A} \sim c$ is the Alfv\'en velocity, and $I_\mathrm{A}$ is the total electric current in the Alfv\'en wing. 
A part $\dot E_\mathrm{J}$ of this power is dissipated into the companion by Joule heating
\begin{equation} \label{puissance_effet_Joule}
\dot E_\mathrm{J} = \frac{I_\mathrm{A}^2}{\sigma_\mathrm{c} h} = \frac{\dot E_\mathrm{A}}{\mu_0 c \sigma_\mathrm{c} h}
\end{equation}
where $\sigma_\mathrm{c}$ is the conductivity of the material constituting the companion, and $h $ is the thickness of the electric current layer. For a small body, relatively to the pulsar wavelength $c/\Omega_*$, we have $ h \sim R_\mathrm{c} $.
For a rocky companion, $\sigma_\mathrm{c} \sim 10^{-3}$ mho.m$^{-1}$, whereas for iron $\sigma_\mathrm{c} \sim 10^{7}$ mho.m$^{-1}$.

Actually, the AW takes its energy from the pulsar wave Poynting flux, and we could consider that $\dot E_\mathrm{A}$ is also a fraction of the power 
$f \dot E_\mathrm{P}$.
We could then conceal this term in the estimate of $f$.  Nevertheless, we keep it explicitly for the estimate of the minimal companion size  
that could trigger a FRB, because we need to know $\dot E_\mathrm{A}$ for the estimation of the FRB power. 

\begin{table*}
	\caption{Notations and abbreviations. The indicated section is their first place of appearance.} 
	\label{table_notation} 
	\centering 
	\begin{tabular}{|l|l|l|}
		\hline 
		Notation  	& Meaning 			& Which section  \\
		\hline	\hline
		NS & neutron star & \\
		AW & Alfv\'en wing &\\
		FRB & fast radio burst & \\
		$RM$ & Faraday rotation measure & \ref{sec_faraday}\\
		\hline
		$R_*, P_*, B_*$ & NS radius, spin period and surface magnetic field & \\
		$\tau_\mathrm{sd}$ & pulsar spin-down age & \\
		$r_\mathrm{LC}$ & light cylinder radius & \\
		$\gamma$ & pulsar wind Lorentz factor &  \ref{section_AW}\\
		$r$ & distance between the NS and its companion & \ref{section_solid_angle}\\
		$T_\mathrm{orb}, n_\mathrm{orb}$ & orbital period and orbital frequency of the companion & \ref{sec_duration_multiplicity}\\
		$R_\mathrm{c}$ & companion radius & \ref{section_AW}\\
		$\sigma_\mathrm{c}, \rho_\mathrm{c}$ & companion conductivity and mass density &\\
		\hline
		$I_A$ & electric current in the Alfv\'en wing &  \ref{section_AW_general}\\
		$S$ & flux density of the FRB radio emission on Earth & \ref{section_AW}\\
		$\epsilon$ & yield of the radio emission & \ref{section_AW}\\
		$D$ & distance between the NS and Earth &  \ref{section_AW}\\
		$\Delta f$ & bandwidth of the FRB radio emissions &  \ref{section_AW} \\
		$f_\mathrm{ce,o} $ & frequency in the observer's frame of electron cyclotron motion in source frame & \ref{sec_freq}\\
		\hline
		$\sigma_\mathrm{source}$ & surface covered by the source & \ref{section_solid_angle} \\
		$\Omega_\mathrm{\sigma}$ & solid angle of the source seen from the NS & \ref{section_solid_angle} \\
		$\Omega_\mathrm{A}$ & solid angle covered by the FRB emission in the source frame & \ref{section_solid_angle} \\
		$\Omega_\mathrm{beam}$ &solid angle covered by the FRB emission in the observer's frame & \ref{section_solid_angle} \\
		$\alpha_{\sigma},\alpha_\mathrm{beam}$ & summit angle of conical solid angles $\Omega_{\sigma}$ and $\Omega_\mathrm{beam}$  & \ref{section_solid_angle} \\
		$\tau_{\rm beam}$ & contribution of the angular opening of the beam to the FRB duration& \ref{burst_duration}\\
		$\tau_{\rm source}$  &contribution of the size of the source to the FRB duration& \ref{burst_duration}\\
		$\tau_{\rm w}$ & typical duration of a bunch of bursts associated with a single asteroid& \ref{sec_discussion_multiplicity}\\
		$\alpha_{\rm w}$ & summit angle of the cone swept by the signal under the effect of wind turbulence & \ref{sec_discussion_multiplicity}\\
		$n_{\rm g}$ & rate of occurence of bunches of FRB associated with a single companion & \ref{sec_how_many_asteroids}\\ 
		$\alpha$ & maximum orbital inclination angle in the asteroid belt & \ref{sec_how_many_asteroids}\\
		$N$ & total number of asteroids & \ref{sec_how_many_asteroids}\\
		\hline
		$\dot E_\mathrm{A}$  & electromagnetic power of the Alfv\'en wing			& \ref{section_AW_general}\\
		$\dot E_\mathrm{iso}, \dot E'_\mathrm{iso} $ &equivalent isotropic luminosity of the source & \ref{section_AW}, \ref{section_AW_bis}\\
		$\dot E_\mathrm{T}$ \& $L_\mathrm{T}$  & 	thermal radiation of the neutron star		&\ref{section_thermal}\\
		$\dot E_\mathrm{P}$  & 	Poynting flux of the pulsar wave				& \ref{section_poynting}\\
		$Q_\mathrm{{abs}}$& Poynting flux absorption rate by the companion & \ref{section_poynting}\\
		$Q_\mathrm{{max}}$& maximal value of $Q_\mathrm{{abs}}$ retained in our analysis & \ref{section_poynting}\\
		$\dot E_\mathrm{NT}$ \& $L_\mathrm{NT}$  & non-thermal photons of the neutron star	&  \ref{section_non_thermal}\\
		$\dot E_\mathrm{W}$ \& $L_\mathrm{W}$  & 	particles in the pulsar wind			& \ref{section_wind_particles}\\
		$\dot E_\mathrm{J}$  & 	Joule heating by the Alfv\'en wing current			& \ref{section_Joule_AW}\\
		$f L_\mathrm{sd}$  & fraction of the pulsar rotational energy loss & \\
		& injected into the pulsar wave Poynting flux					& \ref{section_poynting}\\
		$(1-f) g L_\mathrm{sd}$  & fraction of the pulsar rotational energy loss & \\
		& transformed into wind particle energy and &\\ &non-thermal radiation in the direction of the companion			& \ref{section_wind_and_non_thermal}\\		
		\hline 
	\end{tabular}
\end{table*}

\section{Fluctuations of the dispersion measure} \label{section_fluctuation_DM}

The groups of bursts from FRB 121102 analyzed in \citet{Zhang_2018} show large variations of dispersion measure (DM). With an average value close to $DM$=560 pc.cm$^{-3}$, the range of fluctuations $\Delta DM$ is larger than 100 pc.cm$^{-3}$, sometimes with large variations for bursts separated by less than one minute. Since the variations are strong and fast, we cannot expect perturbations of the mean plasma density over long distances along the line of sight. If real, the origin of these fluctuations must be sought over short distances. If the plasma density perturbation extends over a distance of $10^4$ km, the corresponding density variation reaches $\Delta n \sim 10^{11}$ cm$^{-3}$, i.e. the Goldreich-Julian density near the neutron star surface. Such large fluctuations are very unlikely in the much less dense pulsar wind, where radio wave generation takes place in the present paper. 
Are these large DM variations fatal for our model ?

The DM fluctuations or the (less numerous) bursts from FRB 121102 listed in the FRBcat database are more than one order of magnitude smaller than those reported by \citet{Zhang_2018}. How reliable are the DM estimates in the latter paper ? This question is investigated in \citet{Hessels_2019}. The authors note that the time-frequency sub-structure of the bursts, which is highly variable from burst to burst, biases the automatic determination of the DM. They argue that it is more appropriate to use a DM metrics that maximizes frequency-averaged burst structure than the usual frequency integrated signal-to-noise peak. The DM estimates based on frequency-averaged burst structure reveal a very small dispersion $\Delta DM \le 1$ pc.cm$^{-3}$. They also notice an increase by  1-3 pc.cm$^{-3}$ in 4 years, that is compatible with propagation effects into the interstellar and intergalactic medium. 

The above discussion is likely relevant to other FRB repeaters, such as FRB 180916.J0158+65, since all of them show analogous burst sub-structures. Thus we conclude that the large DM variations observed, for instance in \citet{Zhang_2018}, are apparent and actually related to burst sub-structures. They depend more on the method used to estimate the DM than on real variations of plasma densities. Therefore, they do not  constitute a problem for our pulsar--asteroids model of FRB.



\end{document}